\documentclass[sigconf, nonacm]{acmart}

\newcommand\vldbdoi{XX.XX/XXX.XX}

\usepackage{fancyhdr}
\usepackage[linesnumbered,vlined,ruled]{algorithm2e}
\usepackage{algpseudocode}
\usepackage{graphicx}
\usepackage{textcomp}
\usepackage{xcolor}
\usepackage{bm}
\usepackage{booktabs}
\usepackage{multirow}
\usepackage[listings]{tcolorbox}
\usepackage{enumitem}
\usepackage{microtype}
\usepackage{subfigure}
\usepackage{booktabs} 
\usepackage{amsfonts} 

\usepackage{amssymb}
\usepackage{amsmath}
\usepackage{sansmath}
\usepackage{subfigure}
\usepackage{amsthm}
\usepackage{tcolorbox}
\usepackage{colortbl}
\usepackage{tikz}
\usepackage{url}
\usepackage{pifont}
\usepackage{makecell}
\usepackage{cleveref}
\usepackage{pgfplots}
\usepackage{setspace}
\usepackage[dvipsnames]{xcolor}

\usepackage{listings}
\usepackage{xcolor}

\usepackage{tabularx}
\usepackage{booktabs}
\usepackage{placeins}

\usepackage{array}
\usepackage{multirow}
\usepackage{siunitx}
\usepackage{graphicx}

\definecolor{sqlKeyword}{RGB}{34, 139, 34}      
\definecolor{sqlString}{RGB}{200, 0, 0}     
\definecolor{sqlComment}{RGB}{0, 128, 128}    
\definecolor{sqlArrow}{RGB}{0, 128, 128}      

\crefformat{section}{\S#2#1#3} 
\crefformat{subsection}{\S#2#1#3}
\long\def\comment#1{}

\definecolor{LightCyan}{rgb}{0.88,1,1}
\definecolor{LightRed}{rgb}{1,0.88,1}
\definecolor{LightYellow}{rgb}{1,1,0.88}
\definecolor{LightGray}{gray}{0.8}

\begin{document}

\title{\Sema: A High-performance System for LLM-based Semantic Query Processing}


\author{Kangkang Qi}
\affiliation{%
  \institution{Beijing Institute of Technology}
  \country{}}
\email{qikangkang0126@gmail.com}

\author{Dongyang Xie}
\affiliation{%
  \institution{Wuhan University}
  \country{}}
\email{xdy1122310013@163.com}

\author{Wenbo Li}
\affiliation{%
  \institution{Wuhan University}
  \country{}
}
\email{liwenbo82@whu.edu.cn}

\author{Hao Zhang}
\affiliation{%
 \institution{The Chinese University of Hong Kong}
 \country{}}
\email{zhanghaowuda12@gmail.com}

\author{Yuanyuan Zhu}
\affiliation{%
  \institution{Wuhan University}
  \country{}}
\email{yyzhu@whu.edu.cn}

  \author{Jeffrey Xu Yu}
\affiliation{%
  \institution{HKUST (Guangzhou)}
  \country{}}
  \email{jeffreyxuyu@hkust-gz.edu.cn}

\author{Kangfei Zhao}
\authornote{corresponding author}
\affiliation{%
  \institution{Beijing Institute of Technology}
  \country{}}
\email{zkf1105@gmail.com}

\newcommand{\nthesection}{\arabic{section}}

\newcounter{remark}[section]
\renewcommand{\theremark}{\nthesection.\arabic{remark}}
\newenvironment{remark}{\begin{em}
        \refstepcounter{remark}
        {\vspace{1ex}\noindent\bf Remark \theremark:}}{
        \end{em}\vspace{1ex}} 



\newcommand{\proofsketch}{\noindent{\bf Proof Sketch: }}
\newcommand{\myproof}{\noindent{\bf Proof: }}


\newcommand{\eop}{\hspace*{\fill}\mbox{$\Box$}}
\newcommand{\stitle}[1]{\vspace{1ex} \noindent{\bf #1}}
\newcommand{\etitle}[1]{\vspace{1ex}\noindent{\underline{\em #1}}}
\newcommand{\kw}[1]{{\ensuremath {\mathsf{#1}}}\xspace}
\newcommand{\algocomment}[1]{\footnotesize $\rhd$ \emph{#1}}

\newcommand{\mat}[2]{{\begin{tabbing}\hspace{#1}\=\+\kill #2\end{tabbing}}}
\newcommand{\ls}{\hspace{0.1in}}
\newcommand{\beqn}{\begin{eqnarray*}}
\newcommand{\eeqn}{\end{eqnarray*}}

\newcounter{ccc}
\newcommand{\bcc}{\setcounter{ccc}{1}\theccc.}
\newcommand{\icc}{\addtocounter{ccc}{1}\theccc.}

\newcommand{\tabincell}[2]{\begin{tabular}{@{}#1@{}}#2\end{tabular}} 

\newcommand{\kwnospace}[1]{{\ensuremath {\mathsf{#1}}}}

\newcommand{\radd}[1]{\textcolor{red}{#1}}
\newcommand{\cmark}{\ding{51}}%
\newcommand{\xmark}{\ding{55}}%

\newcommand{\Sema}{\textsc{Sema}\xspace}
\newcommand{\AGM}{\kw{AGM}}
\newcommand{\GHD}{\kw{GHD}}
\newcommand{\ALSS}{\kw{ALSS}}
\newcommand{\LSS}{\kw{LSS}}
\newcommand{\AL}{\kw{AL}}
\newcommand{\MLP}{\kw{MLP}}
\newcommand{\IN}{\kw{IN}}
\newcommand{\LSSFRE}{{$\texttt{LSS-fre}$}\xspace}
\newcommand{\LSSEMB}{{$\texttt{LSS-emb}$}\xspace}
\newcommand{\LSSCON}{{$\texttt{LSS-con}$}\xspace}
\newcommand{\EmptyHeaded}{{\sl EmptyHeaded}\xspace}
\newcommand{\Graphflow}{{\sl Graphflow}\xspace}
\newcommand{\GraphQL}{{\sl GraphQL}\xspace}
\newcommand{\wordtovec}{{\sl word2vec}\xspace}
\newcommand{\ProtBert}{{\sl ProtBert}\xspace}
\newcommand{\DeepWalk}{{\sl DeepWalk}\xspace}
\newcommand{\prone}{{\sl ProNE}\xspace}
\newcommand{\GCN}{{\sl GCN}\xspace}
\newcommand{\GNN}{{\sl GNN}\xspace}
\newcommand{\GIN}{{\sl GIN}\xspace}
\newcommand{\GAT}{{\sl GAT}\xspace}
\newcommand{\SAGE}{{\sl GraphSAGE}\xspace}
\newcommand{\GCARE}{{\sl G-CARE}\xspace}
\newcommand{\WJ}{{$\texttt{WJ}$}\xspace}
\newcommand{\CS}{{$\texttt{CS}$}\xspace}
\newcommand{\CSET}{{$\texttt{CSET}$}\xspace}
\newcommand{\IMPR}{{$\texttt{IMPR}$}\xspace}
\newcommand{\JSUB}{{$\texttt{JSUB}$}\xspace}
\newcommand{\BS}{{$\texttt{BS}$}\xspace}
\newcommand{\SumRDF}{{$\texttt{SumRDF}$}\xspace}
\newcommand{\NRP}{{\sl NRP}\xspace}
\newcommand{\DeepDB}{{\sl DeepDB}\xspace}
\newcommand{\Naru}{{\sl Naru}\xspace}
\newcommand{\DBEst}{{\sl DBEst}\xspace}
\newcommand{\CONF}{{$\texttt{CON}$}\xspace}
\newcommand{\MAR}{{$\texttt{MAR}$}\xspace}
\newcommand{\ENT}{{$\texttt{ENT}$}\xspace}
\newcommand{\CTC}{{$\texttt{CTC}$}\xspace}
\newcommand{\RAND}{{$\texttt{RAN}$}\xspace}
\newcommand{\ENS}{{$\texttt{ENS}$}\xspace}
\newcommand{\GQL}{{$\texttt{GQL}$}\xspace}
\newcommand{\ORI}{{$\texttt{ORI}$}\xspace}
\newcommand{\GFlow}{{$\texttt{GFlow}$}\xspace}

\newcommand{\imdb}{\kw{IMDB}}
\newcommand{\forest}{\kw{forest}}
\newcommand{\dblp}{\kwnospace{DBLP}}
\newcommand{\gene}{\kwnospace{GENE}}
\newcommand{\aids}{\kw{aids}}
\newcommand{\hprd}{\kw{hprd}}
\newcommand{\yeast}{\kw{yeast}}
\newcommand{\wordnet}{\kw{wordnet}}
\newcommand{\youtube}{\kw{youtube}}
\newcommand{\eu}{\kw{eu2005}}
\newcommand{\yago}{\kw{yago}}
\newcommand{\uncertain}{\varphi}
\newcommand{\uncertainLC}{\varphi_{\kw{CON}}}
\newcommand{\uncertainM}{\varphi_{\kw{MAR}}}
\newcommand{\uncertainH}{\varphi_{\kw{ENT}}}
\newcommand{\uncertainCT}{\varphi_{\kw{CTC}}}
\newcommand{\Fagg}{\phi_{{a}}}
\newcommand{\Fcom}{\phi_{{c}}}

\newcommand{\ordemb}{\kw{OrderEmbedding}}
\newcommand{\erm}{\kw{ERM}}

\newcommand{\loss}{\mathcal{L}}
\newcommand{\mse}{\mathcal{L}_{mse}}
\newcommand{\batch}{\mathcal{B}}
\newcommand{\relu}{\mathsf{ReLU}}
\newcommand{\softmax}{\mathsf{softmax}}
\newcommand{\COUNT}{\kw{cnt}}
\newcommand{\AVG}{\kw{avg}}
\newcommand{\SUM}{\kw{sum}}
\newcommand{\MIN}{\kw{min}}
\newcommand{\MAX}{\kw{max}}
\newcommand{\dsb}{\kw{dsb}}
\newcommand{\job}{\kwnospace{JOB}-\kw{light}}

\newcommand{\tbd}{{\color{green}TBD}}

\newcommand{\Model}{\mathcal{M}}
\newcommand{\Dataset}{\mathcal{Q}}
\newcommand{\Prob}{\mathcal{P}}

\newcommand{\Real}{\mathbb{R}}

\newcommand{\Qerror}{\kwnospace{q}\mbox{-}\kw{error}}

\newcommand{\PyTorch}{{\sl PyTorch}\xspace}
\newcommand{\Accelerate}{{\sl Accelerate}\xspace}

\newcommand{\norm}[1]{\left\lVert#1\right\rVert}

\newcommand{\ccross}{\ding{56}}

\newcommand{\SemaSQL}{\textsc{SemaSQL}\xspace}

\newcommand{\SemFilter}{\textit{SemFilter}\xspace}
\newcommand{\SemProj}{\textit{SemProj}\xspace}
\newcommand{\SemJoin}{\textit{SemJoin}\xspace}
\newcommand{\SemAgg}{\textit{SemAgg}\xspace}
\newcommand{\SemOrderBy}{\textit{SemOrderBy}\xspace}

\newcommand{\Lotus}{\text{Lotus}\xspace}
\newcommand{\FlockMTL}{\text{FlockMTL}\xspace}
\newcommand{\Palimpzest}{\text{Palimpzest}\xspace}
\newcommand{\OpenRouter}{{\sl OpenRouter}\xspace}
\newcommand{\MinTime}{{\sl MinTime}\xspace}

\newcommand{\True}{\text{True}\xspace}
\newcommand{\False}{\text{False}\xspace}
\newcommand{\bigo}{\ensuremath{\mathcal{O}}}
\newcommand{\revise}[1]{\textcolor{blue}{#1}}

\renewcommand{\shortauthors}{Kangkang Qi et al.}

\begin{abstract}

The integration of Large Language Models (LLMs) into data analytics has unlocked powerful capabilities for reasoning over bulk structured and unstructured data. However, existing systems typically rely on either DataFrame primitives, which lack the efficient execution infrastructure of modern DBMSs, or SQL User-Defined Functions (UDFs), which isolate semantic logic from the query optimizer and burden users with implementation complexities.
The LLM-powered semantic operators also bring new challenges due to the high cost and non-deterministic nature of LLM invocation, where conventional optimization rules and cost models are inapplicable for their optimization. 

To bridge these gaps, we present \Sema, a high-performance semantic query engine built on DuckDB that treats LLM-powered semantic operators as first-class citizens. \Sema introduces \SemaSQL, a declarative dialect that allows users seamlessly inject natural language expressions into standard SQL clauses, enabling end-to-end optimization and execution. 
At the logical level, the optimizer of \Sema compresses natural language expressions and deduces relational constraints from semantic operators.
At runtime, \Sema employs Adaptive Query Execution (AQE) to dynamically reorder operators, fuse semantic operations, and apply prompt batching. This approach seeks a Pareto-optimal execution path balancing token consumption and latency under accuracy constraints. 
We evaluate \Sema on 20 semantic queries across classification, summarization, and extraction tasks. Experimental results demonstrate that \Sema achieves $2-10 \times$ speedup against three baseline systems while achieving competitive result quality. 

\end{abstract}





\maketitle



\section{Introduction}
The powerful semantic reasoning capabilities of modern language models (LLMs) are unlocking new opportunities for AI-driven analytics over both structured and unstructured data. Applications in science, healthcare, business and industry increasingly require complex reasoning over large datasets, e.g., summarizing recent research, extracting biomedical insights from patient records, or analyzing corporate meeting transcripts at scale. 
To realize these capabilities, analytics systems need to orchestrate sophisticated, bulk semantic processing with efficiency and usability, promising a revolutionize on existing database systems. 
This vision raises two fundamental questions: 
\emph{How should developers express semantic queries? And how should developers design a data analytics system to achieve high efficiency and accuracy?}

Recent systems have begun to explore LLM-powered semantic query processing for both structured~\cite{DBLP:journals/corr/abs-2504-01157, liu2025optimizing} and unstructured data~\cite{DBLP:conf/icde/WangF25, DBLP:journals/corr/abs-2410-12189, DBLP:journals/corr/abs-2409-00847}. 
A variety of systems DocETL~\cite{DBLP:journals/corr/abs-2410-12189},  
ZenDB~\cite{DBLP:journals/corr/abs-2405-04674}, DocWanger~\cite{DBLP:journals/corr/abs-2504-14764}, 
Aryn~\cite{DBLP:journals/corr/abs-2409-00847} and Unify~\cite{DBLP:conf/icde/WangF25} support semantic queries expressed in extended SQL, declarative operations, or natural language. These systems employ LLMs not only to interpret queries, but also to optimize execution through techniques like query planning~\cite{DBLP:journals/corr/abs-2409-00847}, rewriting, prompt refinement, and operation decomposition~\cite{DBLP:conf/icde/WangF25}. 
Palimpzest~\cite{liu2025palimpzest} and its optimizer Abacus~\cite{russo2025abacus} focus on reducing LLM query latency and cost through logical and physical optimizations, including Pareto-efficient planning. Lotus~\cite{patel2024lotus} formulates LLM-based semantic operations for tabular data and proposes efficient execution algorithms for costly operators. 
Recent benchmarks~\cite{akillioglu2025, DBLP:journals/corr/abs-2504-01157} have highlighted key challenges in supporting semantic query processing in existing databases, such as ensuring structured outputs, enabling batch processing, and effective query planning.

Current systems typically express semantic queries in two different ways: DataFrame primitives~\cite{liu2025palimpzest, russo2025abacus, patel2024lotus}  and SQL user-defined functions (UDFs)~\cite{DBLP:journals/corr/abs-2504-01157, akillioglu2025, liu2025optimizing}, each highly intertwined with the underlying system design. 
Systems built on DataFrame primitives define custom semantic operations and optimize them with specialized algorithms, but often lack the efficient execution infrastructure of a DBMS. This forces developers to redeploy execution environments such as vectorized and pipeline parallel execution, I/O and memory management. 
On the other hand, integrating semantic operators as SQL UDFs offers greater flexibility, but places the burden of implementing, debugging, and maintaining complex AI-based operators on users, raising concerns about security, reliability, and efficiency.

With the growing demand for semantic queries, a core set of semantic operators, such as semantic filter, projection, join, and aggregation, has emerged as an AI-powered analogy to classic relational operators~\cite{DBLP:conf/icde/WangF25, patel2024lotus}. 
A natural envision arising from this trend is \emph{whether we can efficiently support these semantic operators as native database operators, thereby fully leveraging DBMS optimization processing?}
However, fulfilling this envision is not merely about implementation add-ons. 
Unlike relational operators, which have deterministic semantics, stable algebraic laws, and cost models approximated from data statistics, semantic operators are context- and model-dependent: their execution depends not only on input tuples, but also on the natural language (NL) expression, prompting strategy, and the specific LLM being invoked. 
LLM inference optimizations in external LLM serving environments, such as KV cache, Paged Attention, and continuous batching also impact the execution performance significantly. Modeling the cost of these LLM-powered operators also beyond the modeling of complex, CPU-expensive operators~\cite{hellerstein1993predicate}. 
This shift breaks key assumptions behind traditional query optimization: standard rewrite rules may be unsafe, selectivity and cost are unable to estimate statically. 
These challenges call for new system-level optimizations tailored for semantic operators. 
The system should treat semantic operators as first-class citizens and focus on reducing the cost of LLM invocations, which dominate the performance of query evaluation. Furthermore, semantic operators also bring additional concerns of multiple optimization objectives, including query accuracy, latency, and the monetary cost of LLM calls.

\begin{figure}
    \centering
    \includegraphics[width=0.95\linewidth]{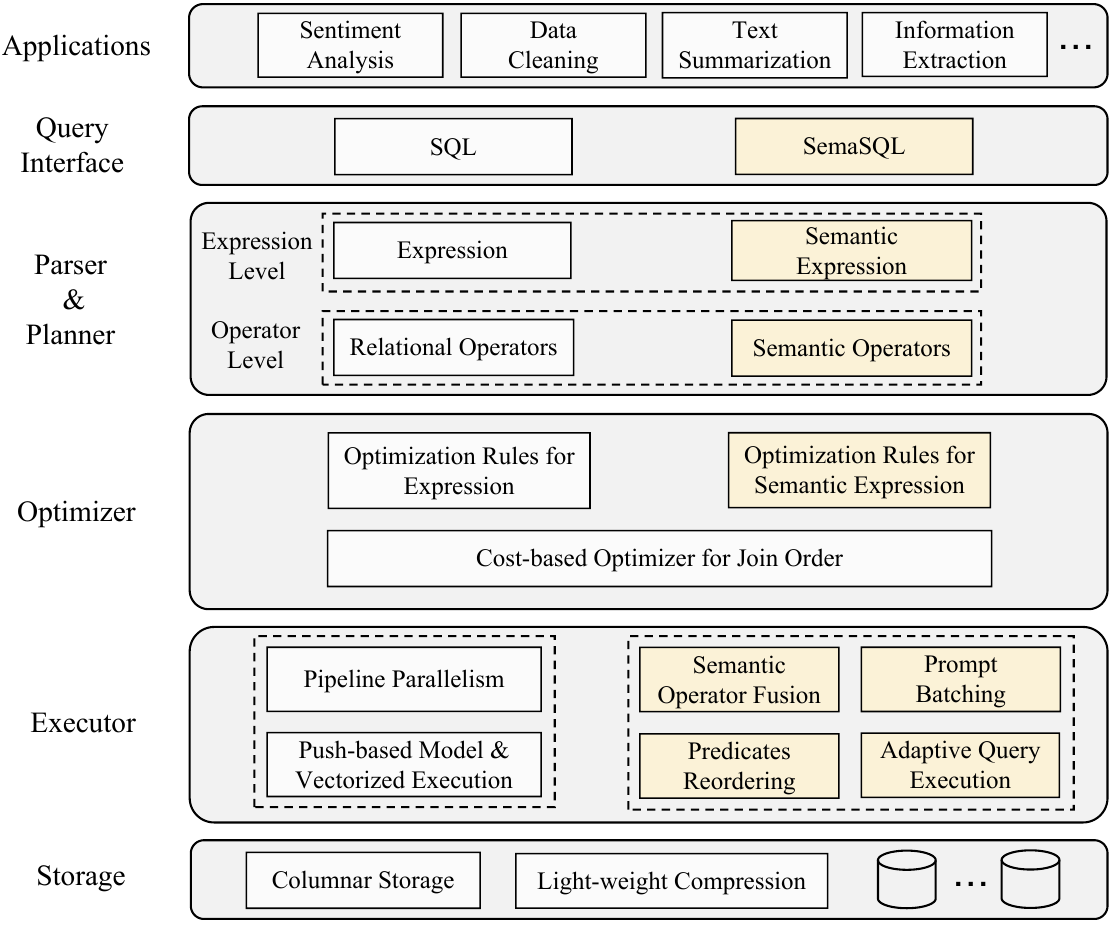}
    \vspace{-2ex}
    \caption{System architecture of \Sema}
    \label{fig:architecture}
\end{figure}
As a tentative realization, we design and develop a prototype system dubbed \Sema back ended DuckDB~\cite{DBLP:conf/sigmod/RaasveldtM19}, injecting native semantic operators that invoke local or remote LLMs. 
On the basis of vectorized and pipeline execution infrastructure, the design principle of \Sema focuses on reducing the cost of LLM invocations.
Fig.~\ref{fig:architecture} illustrates the architecture of \Sema, which extends the query language interface, parser and planner, optimizer and executor of DuckDB accordingly, as highlighted in Fig.~\ref{fig:architecture}.
To express semantic operators, \Sema introduces \SemaSQL that allows users to insert Natural Language (NL) expressions into SQL clauses. These NL expressions, wrapping DB column names, convey user intent on semantic predicates, ranking, or aggregate criteria. 
To process these native semantic operators efficiently and flexibly, we propose logical intra-operator optimizations in the optimizer for compressing the NL expression and deducing relational operators from semantic operators, where the optimizer leverages auxiliary LLMs to fulfill these optimizations.
In the executor, \Sema integrates semantic operator fusion and prompt batching to reduce the number of LLM invocations in bulk data processing. 
As the statistical patterns of semantic operator instances are hard to depict before execution, \Sema adopts adaptive query execution (AQE) to reorder consecutive semantic operators, together with operator fusion and batch prompting. The AQE attempts to find a Pareto-optimal execution path regarding LLM token consumption and query latency, under the constraint of query accuracy. 
We comprehensively evaluate \Sema on 20 semantic queries spanning 3 analytical tasks (classification, summarization, and keyword extraction) from BIRD~\cite{DBLP:conf/nips/BIRD-Benchmark}. The results demonstrate that \Sema achieves a 2–4× speedup over the DataFrame-based system \Lotus and a 2–10× speedup over pipeline-execution system \Palimpzest, while maintaining competitive result quality compared with both systems and significantly outperforming the UDF-based system \FlockMTL.
The contributions of this paper are summarized as follows. 
\begin{itemize}[leftmargin=*]
    \item We design and develop \Sema, a high-performance semantic query engine built on DuckDB that treats LLM-powered semantic operators as first-class query plan operators, enabling end-to-end semantic query optimization and execution in a columnar and  vectorized database system.
    \item We propose \SemaSQL, a declarative SQL dialect that allows users to inject natural language expressions into standard SQL clauses to define semantic operators, enabling seamless composition of semantic and relational computation.
    \item We introduce optimization techniques tailored to semantic operators to reduce the overhead of LLM invocations collaboratively, including logical-level semantic expression optimizations and runtime optimizations via adaptive query execution (AQE) that combines operator fusion and prompt batching. 
    \item We design a semantic query benchmark consisting of 20 queries over 9 datasets and conduct a comprehensive evaluation of \Sema. Results demonstrate that \Sema significantly outperforms both DataFrame-based and UDF-based systems in query efficiency with competitive result quality, demonstrating high usability.
\end{itemize}
\stitle{Roadmap.} The following of this paper is organized as follows. We introduce the background and our motivation in \cref{sec:preliminary}. \cref{sec:overview} introduces the system overview of \Sema, followed by the details of its optimizer and executor in \cref{sec:optimizer} and \cref{sec:executor}, respectively. \cref{sec:engineering_details} elaborates on critical engineering details of \Sema. 
We report our system evaluation in \cref{sec:exp} and review related works in \cref{sec:related}. We conclude the paper in \cref{sec:conclusion}.

\section{Background}
\label{sec:preliminary}
\subsection{Semantic Operators}

\begin{table*}[!t]
\centering
\footnotesize
\setlength{\tabcolsep}{4pt} 
\caption{Semantic Operators}
\vspace{-2ex}
\label{table:semantic_operators}
\begin{tabular}{c c c r}
\toprule
\textbf{Operator} &\textbf{Expression} & \textbf{Description} & \textbf{\# LLM calls} \\
\midrule
Semantic Filter & $\{t \mid t \in T \wedge M(t, e)= \True\}$ & Select tuples that can pass a NL predicate $e$ & $\bigo(|T|)$  \\
Semantic Projection & $\{M(t, e) \mid t \in T\}$ & Project attribute values specified by a NL predicate $e$ & $\bigo(|T|)$ \\
Semantic Join & $\{ (t_i, t_j) \mid t_i \in T_1 \wedge t_j \in T_2 \wedge M(t_i, t_j, e) =\True \}$ & Join two tables on a join key that pass a NL predicate $e$ & $\bigo(|T_1| \cdot |T_2|)$ \\
Semantic Order By & $[t_1, \cdots, t_n], s.t. \forall [t_i, t_j] \Rightarrow M(t_i, t_j, e) = t_i\preceq t_j$ & Reorder tuples following a NL-specified criteria & $\bigo(|T|^2)$ \\
Semantic Aggregate & $M(\{t_1, \cdots, t_n\}, e), t_1, \cdots,t_n \in T$ & Aggregate tuples according to a NL-specified aggregate function & $\bigo(1)$ \\
\bottomrule
\end{tabular}
\end{table*}

A semantic operator is a declarative transformation on tabular data, which is parametrized by a natural language expression as a predicate. In this paper, we consider semantic operators powered by Large Language Models (LLMs). We use $T(A_1, \cdots, A_j)$ or $T$ for short to denote a table, where $A_1, \cdots, A_j$ are the attributes of $T$. $t \in T$ denotes a tuple in Table $T$ and $t(A_j)$ denotes the value of attribute $A_j$ in tuple $t$. $M: \mathcal{T} \mapsto \mathcal{T}$ is a pre-trained LLM that accepts input and generates output in textual space $\mathcal{T}$.
\Sema starts from a relational algebra generalization on semantic query processing, a coreset of semantic operators in Lotus~\cite{patel2024lotus}, while other specific systems may provide additional operators beyond these operators.

Table~\ref{table:semantic_operators} summarizes these five semantic operators we implemented in \Sema, i.e., Semantic Filter (\SemFilter), Semantic Projection (\SemProj), Semantic Join (\SemJoin), Semantic OrderBy (\SemOrderBy), and Semantic Aggregate (\SemAgg). 
Each operator is implemented by a default reference algorithm.
For \SemFilter and \SemProj, their reference algorithms scan the table and invoke an LLM to evaluate each tuple in table $T$, and the LLM is requested to return \True or \False for \SemFilter or textual attribute for \SemProj, respectively, 
which take $\bigo(|T|)$ LLM invocations.
For \SemJoin, the reference algorithm performs a nested loop join on two tables $T_1$ and $T_2$, and uses an LLM to evaluate whether a pair of tuples can be joined on a semantic predicate, which takes $\bigo(|T_1|\cdot|T_2|)$ LLM invocations. 
The reference algorithm of \SemOrderBy is a Selection Sort algorithm that compares two pairs of tuples in $T$ regarding a semantic comparison criteria, resulting in $\bigo(|T|^2)$ LLM comparisons.
For \SemAgg, the reference algorithm performs a many-to-one reduce that puts all the values of an attribute in $T$ into an LLM prompt and generates the aggregated result. 
Managing finite context limits of the underlying LLM is implemented by the system, where hierarchical aggregation over subsets of values can be performed.
Similar to relational operators, the evaluation of one semantic operator can be implemented by different algorithms; however, supporting LLM-based semantic operators inside a database engine is non-trivial, not simply an implementation.

\subsection{Semantic Operators Handling}

Unlike relational operators, whose behavior is grounded in deterministic and decidable semantics, well-defined algebraic laws, and stable cost models, semantic operators are inherently context- and model-dependent. Their outcomes depend not only on input tuples, but also on the NL expression, prompting strategy, and the specific LLM invoked. As a result, evaluation is intrinsically uncertain and cannot be characterized by a single canonical truth function.
This shift in semantics also weakens the foundations that classical query optimizers rely on. Standard algebraic rewrite rules may not hold: for example, \SemFilter is commutative only under restrictive conditions (e.g., tuple-independent prompting and zero-temperature inference), while associativity does not generally apply. For \SemProj and \SemJoin, commutativity and associativity are typically not semantics-preserving, which makes traditional rule-based rewriting unsafe or inapplicable. Finally, the cost profile of semantic operators is dominated by external inference environments, i.e., latency, throughput, and monetary cost vary with models, prompts, inference optimization, and caching mechanism; therefore, they are difficult to estimate from database statistics alone. Together, these properties call for a system design that treats semantic operators as first-class citizens rather than black-box add-ons.

Existing solutions for semantic query processing are constrained by their execution environments. Implementations built on lightweight libraries (e.g., Pandas or custom DataFrame APIs) provide accessible programming models but lack mature database capabilities such as vectorized execution, pipeline parallelism, and robust query optimization, resulting in inefficient processing for large-scale hybrid analytical workloads. 
In contrast, database-centric approaches frequently encapsulate LLM calls as user-defined functions (UDFs). While convenient, this makes semantic operators opaque to the optimizer and executor, preventing system-level optimizations tailored to semantic workloads.

Motivated by these gaps, we built \Sema, a prototype system that integrates semantic operators directly into DuckDB to process semantic operators at scale by exploiting a high-performance OLAP execution engine with an expression-centric design, vectorized processing, and pipeline parallelism, capabilities that DuckDB developed.
By embedding semantic primitives into the query planner, optimizer, and executor, \Sema provides a concrete platform for exploring semantic-specific optimization, i.e., expression optimization and adaptive query execution, aiming to reduce the overhead of LLM invocations. 
In this work, we focus on the system design and end-to-end integration of semantic operators in a relational engine. We do not aim to fully optimize each individual operator or cover the full space of production concerns. We leave the following to future work: specialized algorithmic improvements for individual operators, storage-level optimizations, LLM routing and model selection, and comprehensive mechanisms for privacy, governance, and fault tolerance in external inference.

\section{System Overview}
\label{sec:overview}
We present an overview of \Sema, including its query interface (\cref{sec:overview:interface}) and workflow (\cref{sec:overview:workflow}). Fig.~\ref{fig:architecture} shows the system architecture of \Sema.

\subsection{The Query Syntax}
\label{sec:overview:interface}
\Sema introduces \SemaSQL, a declarative syntax with injection of Natural Language (NL) expression, for semantic query processing. The design of \SemaSQL follows three main principles:
\emph{Transparency}: isolation from underlying implementation and optimization, 
\emph{Flexibility}: sufficient expressive capability of user intent with high utility, \emph{Compatibility}: extension of \SemaSQL compatible with the syntax standard and implementation of SQL. 

The first-class citizen of \SemaSQL is an NL expression with one or multiple placeholders, denoted as $s  `\cdots \{ \} \cdots\text{'}$ where $\{ \}$ is a placeholder that identifies a mention of columns/attributes or the NL description of the attributes in the expression, instructing the system to process the corresponding tables and attributes.
Semantic operators are implicitly implemented by injecting NL expressions into SQL clauses such as \texttt{SELECT}, \texttt{WHERE}, \texttt{JOIN ON}, etc. 
Specifically, \SemProj injects an NL expression into \texttt{SELECT} and \SemFilter injects an NL expression into \texttt{WHERE} or \texttt{HAVING} clauses, respectively.
For \SemJoin, we inject an NL expression into a \texttt{JOIN ON} clause, which is also compatible with other join types, including \texttt{OUTER JOIN}, \texttt{SEMI JOIN}, etc. 
For \SemAgg, we introduce a special aggregation function \texttt{sem\_agg()} that wraps the NL expression, and \texttt{sem\_agg()} can be used together with the standard \texttt{GROUP BY} clause.
In addition, users can use \texttt{AS} to rename new textual columns, e.g., the columns generated from \SemProj and \SemAgg, and these columns can be further mentioned in other NL expressions. 
Fig.~\ref{fig:sem_query} demonstrates a \SemaSQL query containing two predicates of \SemFilter in the \texttt{WHERE} clauses. 
The query performs a relational join on two tables, with two relational operators and two \SemFilter operators on the {\texttt{translated\_review}} attribute of table {\texttt{user\_reviews}}. Here, NL expressions are used as the semantic predicates whose valuations are inferred by an LLM. 
Additional exemplar queries can be found in Appendix~\ref{sec:appendix:example}, Fig.~\ref{fig:semantic_projection}-\ref{fig:semantic_join}.

\SemaSQL uses interpolation of NL expressions to support semantic operators while preserving the nature of the symbolic logic of SQL. \Sema also supports complex syntax with \SemaSQL such as subqueries, Common Table Expressions (CTEs), and window functions, etc. Unlike NL2SQL, \SemaSQL makes minimal extensions, i.e., the NL expression on SQL, enabling full compatibility with RDBMS and sufficient semantic reasoning.

\subsection{System Workflow}
\label{sec:overview:workflow}
\begin{figure*}[t]
    \centering
    \includegraphics[width=0.95\linewidth]{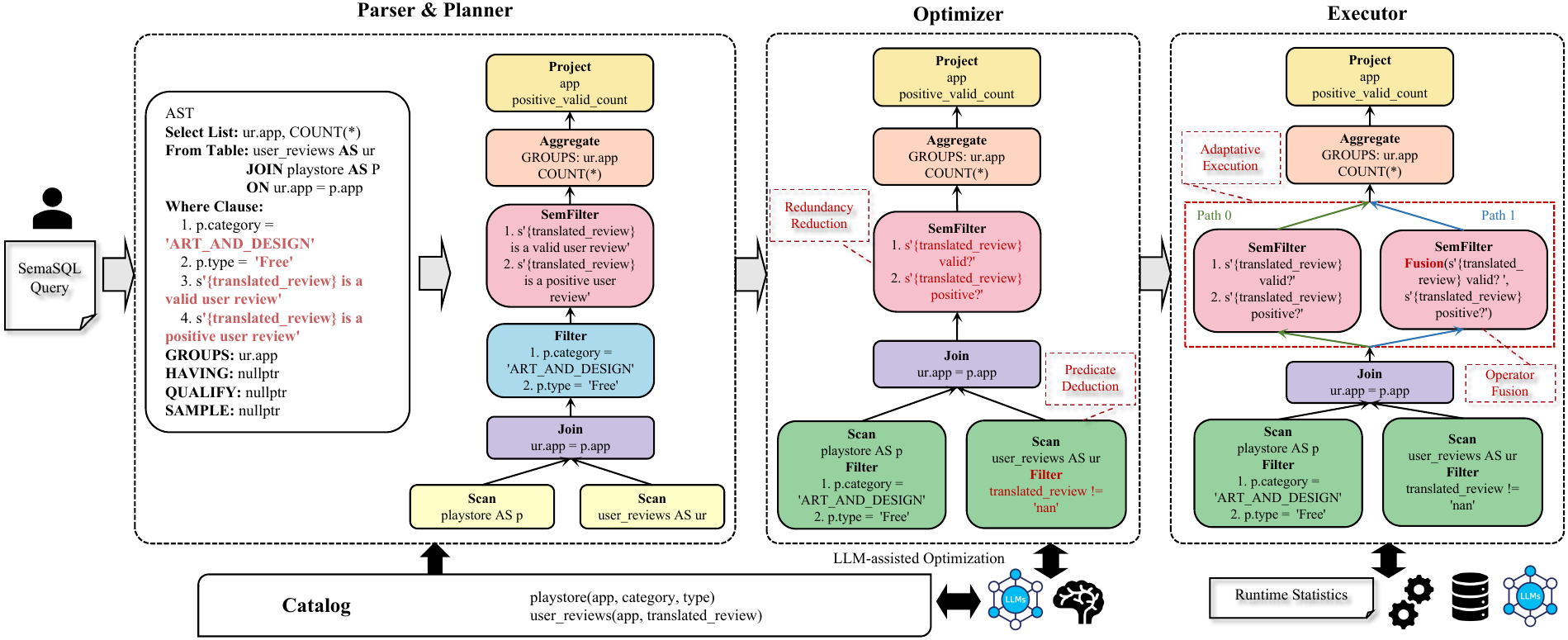}
    \caption{System workflow of \Sema}
    \vspace{-2ex}
    \label{fig:workflow}
\end{figure*}

We use the running example in Fig.~\ref{fig:sem_query} to illustrate the end-to-end workflow of \Sema, which is depicted in Fig.~\ref{fig:workflow}.
The query counts, for each \kw{app},  the reviews that satisfy two semantic predicates  (`is a valid user review' and `is a positive user review'), together with standard relational predicates on \kw{category} and \kw{type}. 
When the user submits the query, \Sema parses \SemaSQL and produces an abstract syntax tree (AST) and an initial logical plan in which NL expressions are identified as explicit semantic operators. Here, two \SemFilter operators become first-class plan nodes that \Sema can analyze and transform at the logical and physical levels, respectively.

First, logical NL expression optimization in the optimizer operates before conventional SQL rewrites and focuses on reducing and restructuring intra-operator semantic expressions. 
Specifically, \Sema applies redundancy-aware expression compression to shorten NL predicates and stabilize prompting. 
More importantly, \Sema performs predicate deduction for \SemFilter: it opportunistically extracts fully or partially deductible symbolic conditions (e.g.,  `\kw{translated\_review} != ``nan''') from an NL predicate, rewrites them into relational filters which are pushed down to pruned tuples early, thereby avoiding unnecessary LLM invocations. These expression optimizations are guided by an auxiliary LLM with CoT prompting. \cref{sec:optimizer} will elaborate on these optimizations injected as special rules for semantic expression in the optimizer. 

Then, physical execution optimization and adaptive query execution (AQE) target the dominant runtime cost of semantic operators, under a vectorized, pipeline-parallel execution engine. 
The executor of \Sema applies semantic operator fusion and prompt batching to reduce the number of LLM calls and improve throughput.
Because the effects of these optimizations and the selectivities of \SemFilter are hard to predict statically, \Sema introduces AQE for \SemFilter operators.  At runtime, the executor micro-executes a small fraction of the input to collect predicate statistics, explores multiple candidate execution paths, and selects a path that best trades off latency and token cost subject to an accuracy tolerance. The selected path is then used to process the remaining data at scale. 
We will present these technical details in \cref{sec:executor}. 

\begin{figure}
\begin{lstlisting}{sql}
-- Calculate the counts of valid and positive reviews for each app of 'Free' type and 'ART_AND_DESIGN' category.
SELECT ur.app, COUNT(*) AS positive_and_valid_count
FROM user_reviews AS ur INNER JOIN playstore AS p
ON ur.app = p.app 
WHERE p.category = 'ART_AND_DESIGN' AND p.type = 'Free'
    AND s'{translated_review} is a valid user review'
    AND s'{translated_review} is a positive user review'
GROUP BY ur.app;
\end{lstlisting}
\vspace{-2ex}
\caption{A running semantic query example}
\label{fig:sem_query}
\end{figure}

\section{The Optimizer of \Sema}
\label{sec:optimizer}
Aligning with the goal of our system design, \Sema's optimizer concentrates on reducing the cost of LLM invocations without compromising accuracy. 
As end users may deliver arbitrary NL expressions in the semantic operators with \SemaSQL, NL expressions are inherently verbose or ambiguous, and their evaluation will bring redundant overhead for LLM reasoning. 
To resolve these issues, \Sema's optimizer introduces specific rules for optimizing these NL expressions, including expression compression and predicate deduction, which are applied to the query plan step by step.
Since NL expressions are free text, traditional rule matching cannot comprehensively capture the compression and deduction patterns, therefore, we use LLM-assisted optimizations.

\stitle{NL Expression Compression.} 
For the NL expression in each semantic operator, the optimizer first employs a light-weight LLM with Chain-of-Thought (CoT) prompting~\cite{DBLP:conf/nips/Wei0SBIXCLZ22} with demonstrations to compress and refine the NL expression. 
The CoT prompting includes decomposed instructions on removing stop words, simplifying sentence structure, e.g., passive to active voice, and consolidating repeated pronoun references to the placeholder, which are listed as multiple rules.
We also inject demonstrations for each rule to illustrate its utility. 
This step introduces only one LLM call for each semantic operator, while potentially reducing LLM inference cost and improving semantic clarity.

\stitle{Predicate Deduction.}
After expression compression, the optimizer seeks the opportunity to deduce semantic filters into relational filters, i.e., rewriting entire or partial NL expressions of \SemFilter as SQL predicates. An entire deduction bypasses LLM invocations, while a partial deduction only surgically replaces partial filter predicates in compound logic with SQL predicates as a necessary condition,  preserving the remaining semantics as more concise NL expressions. This partial deduction enables potentially fewer LLM invocations, with the deducted SQL predicates pushdown.
Predicate Deduction for \SemFilter is analogous to predicate simplification for relational query processing, and the similar idea has also been introduced into ML-based predicates~\cite{chaudhuri2002efficient}.
It is worth mentioning that existing NL2SQL approaches~\cite{deng-etal-2022-recent-advances-in-text-to-sql} cannot directly perform this task. Rule-based approaches fully rely on rule
matchings for SQL translation, which do not possess
flexible reasoning ability for semantic expression. 
DL-based approaches rely on high-quality annotated data for training. LLM-based approaches have the potential to deduce the NL predicates; however, they cannot support partial deduction, i.e., translating a SQL predicate as a necessary condition, with a corresponding verification mechanism.

\Sema supports deducing predicates of categorical matching, numeric comparison, and \kw{LIKE} operations on strings.
First, we use CoT prompting, feeding the DB schema into the context, to guide an LLM to opportunistically identify deductible predicates and then transform them into SQL expressions in string format.
The instruction is explicitly decomposed into multiple steps, starting from extracting relevant tables and columns from the expression, determining if the query possesses objective intent, e.g., numeric or string comparison, and whether it can be mapped to a pattern of symbolic comparison, subsequently judging deductibility, and finally generating a SQL statement. In each step, we provide few-shot examples, which also demonstrate special cases of case variations, fuzzy matches, and data type consistency.
To alleviate LLM hallucination, we equip a back-forward self-reflection  mechanism. Once a syntactically valid SQL statement is deduced, we request the LLM by another CoT prompt to judge whether the SQL statement is a necessary condition for the original NL expression. 
If the deduction of the self-reflection fails, the system falls back to the original \SemFilter, preserving correctness.

\begin{example}(Expression Compression and Predicate Deduction) We use the query in Fig.~\ref{fig:sem_query}, which contains two \SemFilter operators on the attribute  \kw{translated\_view} in table \kw{user\_reviews}. The parser of \Sema transforms the query into an initial unoptimized plan shown as Fig.~\ref{fig:workflow}. In the optimizer, the two NL expressions are first compressed, forming two shorter sentences, respectively. As the placeholders already wrap exact matches of column names, meta-entity resolution directly binds them to the column objects. Subsequently, predicate deduction extracts a SQL predicate `translated\_review != ``nan''' from the \SemFilter with the expression `s\{translated\_review\} is a valid user review'.
Here, an auxiliary LLM is prompted with the DB metadata as context, which indicates that for column \kw{translated\_review}, around 41.8\% of entries are `nan'. The `nan' is obviously an invalid review, but it may not be noticed by the users. Predicate deduction automatically identifies this symbolic logic from a \SemFilter, and the deducted SQL predicate is further pushed down to enable early filtering, as shown in Fig.~\ref{fig:workflow}, bypassing a large number of LLM invocations.
\end{example}

In the optimizer of \Sema, LLM-assisted expression optimizations are applied before rule-based optimization for standard SQL.
By strategically employing LLM-assisted expression optimizations, the optimizer improves the usability of \SemaSQL for general users to access their analytical tasks and enhances the reliability, accuracy, and efficiency of \Sema. We provide the algorithm in the appendix to present the overall procedure of expression optimizations, together with the prompts for predicate deduction by \Sema.

\section{Execution Optimization}
\label{sec:executor}
The executor of \Sema integrates new physical optimizations (\cref{sec:executor:physical}) and Adaptive Query Execution (\cref{sec:aqe}), which are specifically designed for semantic operators.

\subsection{Physical optimization}
\label{sec:executor:physical}
\stitle{Semantic Operator Fusion.}
In semantic query evaluation, each semantic operator typically triggers an individual LLM invocation. While this preserves modularity and operator independence, for complex queries with cascaded semantic operators, intermediate data will be repeatedly materialized and transferred between multiple operators, leading to a high memory footprint, token consumption, and API call latency. 
Inspired by operator fusion for relational operators~\cite{DBLP:journals/pvldb/Neumann11} and linear algebra operators~\cite{DBLP:journals/pvldb/BoehmRHSEP18}, \Sema introduces semantic operator fusion, which combines consecutive semantic operators into one operator.
Unlike fusing conventional operators which synthesizes generated code by compiling optimizations, fusing semantic operators is essentially combining their semantic NL expressions into one expression, with one LLM inference per tuple. 

\Sema supports four patterns of operator fusion as listed in Table~\ref{table:opt_fusion}, between two `unary' semantic operators \SemFilter and \SemProj on one table in one query pipeline. We bootstrap semantic operator fusion from these four patterns, since the fusion of more operators that includes both unary and binary operators entails parsing and combining complex semantics to guarantee the evaluation accuracy. For \SemFilter-first operators, the NL expressions of the two operators to be fused should refer to a common attribute in their placeholders. For \SemProj-first operators, the rename operator \texttt{AS} indicates the intermediate column, which is referred by the expression of the subsequent operator.
Once \Sema identifies a match of these patterns, it concatenates the two NL expressions of the two consecutive operators into a single NL expression and adds an instruction for step-by-step processing.

As Table~\ref{table:opt_fusion} summarizes, a fusion will save $|T|$ LLM invocations for \SemProj-first patterns and $s\cdot |T|$ LLM invocations for \SemFilter-first patterns, respectively, where $s$ is the selectivity of the first \SemFilter in table $T$.

\begin{table}[!t]
\begin{center}
\caption{Operator Fusions in \Sema}
\label{table:opt_fusion}
\footnotesize
\begin{tabular}{c cccc}
\toprule
\# LLM calls &  $\sigma_a \oplus \sigma_b$ & $\Pi_a \oplus \Pi_b$ &  $\sigma_a \oplus \Pi_b$  & $\Pi_a \oplus \sigma_b$ \\
\midrule
Original & $2|T|$ & $2|T|$ & $(1 + s)\cdot |T|$ & $(1 + s)\cdot |T|$ \\
Fused &   $|T|$ & $|T|$  & $|T|$ & $|T|$ \\
\bottomrule
\end{tabular}
\end{center}
\end{table}

\begin{example}(Semantic Operator Fusion) We use the query in Fig.~\ref{fig:sem_query} to illustrate operator fusion, which contains two \SemFilter referring a same column \kw{translated\_review} in table \kw{user\_review}. \Sema  combines the two compressed expressions s'{translated\_review} is valid' and s'{translated\_review} is positive' into a single expression `s{translated\_review} is valid and positive', and then generates a new \SemFilter for the expression to replace the original two operators.
\end{example}

\stitle{Prompt Batching.}
As an analytical system, the vectorized execution of DuckDB is the cornerstone of batch query processing of \Sema. 
On this basis, \Sema further considers improving token utilization and throughput of semantic operators with prompt batching, i.e., batching multiple tuples into a single LLM prompt~\cite{lin2024batchprompt}, for \SemFilter, \SemProj, and \SemJoin.

Despite the reduction of LLM calls, operator fusion and prompt batching break the operator-independent and tuple-independent execution.
A potential risk is that they usually lead to inconsistent inference results compared to one-by-one LLM invocations. 
Taking this issue into account, first, for prompt batching, the executor exploits the structural output of LLM API calls that specify the return value of an API call as a
Json Array. This enforces the output of one batch prompting to generate a consistent number of results. 
Second, when fusing \SemFilter-first patterns, \Sema adds an instruction in the combined prompt to request the LLM explicitly generate intermediate results of the first operator, in order to maintain the evaluation consistency of the fused operator with the original operator flow.
\Sema restricts operator fusion to only two operators, leaving fusion on more operators as further work. 
Third, for \SemFilter, \Sema allows users to specify the tolerance regarding the consistency of results. This user preference will be used to choose the real execution plan by adaptive query execution at runtime, which we will elaborate on in \cref{sec:aqe}.

\subsection{Adaptive Query Execution for \SemFilter}
\label{sec:aqe}

The behavior of semantic operators depends on both the NL expression and the particular LLM used to interpret it, making key planning quantities, such as selectivity, latency, and monetary/token cost difficult to predict statically. Moreover, optimizations that reduce LLM calls (e.g., operator fusion and batch prompting) can introduce hard-to-model accuracy shifts.
To address these challenges, we postpone inter-operator optimizations such as operator reordering and fusion to runtime via an adaptive query execution (AQE) strategy. 
Specifically, the executor of \Sema collects runtime feedback from evaluating a small fraction of data on multiple candidate physical plans, a.k.a., execution paths, and selects an optimal path to process the remaining data. 
Different from relational query processing where latency is typically the unique optimization goal, semantic query processing entails considering multiple objectives including query accuracy, query latency, and monetary cost of LLM invocations. 
Thereby, the AQE of \Sema is designed to find the Pareto optimal path for both query latency and monetary cost, under the constraint of a user-specific tolerance to accuracy inconsistency. If the accuracy loss is more tolerant, or the latency and cost are highly-concerned factors, the system can be easily extended regarding the objective of Pareto optimization.
As the ground-truth of semantic query is usually unavailable beforehand, we use a reference path, e.g., the path generated by the optimizer, without operator fusion and batch prompting, to evaluate the accuracy deviation of other candidate paths.

The AQE framework operates in three phases: the expression exploration phase, the path exploration phase and the path exploitation phase.
Algorithm~\ref{alg:aqe-main} presents the overall procedure of AQE. 
In expression exploration phase (line~\ref{line:phase1:start}-\ref{line:phase1:end}), we use a small fraction of $\delta_1$ tuples to collect statistics of each \SemFilter and the sudo ground-truth of reference path $p_{\text{ref}}$.
Subsequently, we generate a set of candidate paths $\mathcal{P}$ using these statistics, and evaluate the performance of these paths on another small fraction of $\delta_1$ tuples, selecting the optimal paths $p^\star$ in $\mathcal{P}$ regarding either time-efficiency or token-efficiency preference (line~\ref{line:phase2:start}-\ref{line:phase2:end}).
Finally, in the path exploitation phase (line~\ref{line:phase3}), the executor will process the remaining $1- (\delta_1 + \delta_2)$ tuples.
Users can balance the exploration and exploitation via tuning the system parameters $\delta_1$ and $\delta_2$.
Note the overall query results are the concatenation of the results from the three phases: (1) evaluating $\delta_1$ tuples individually on each \SemFilter and intersecting their results,  (2) evaluating  $\delta_2$ tuples by $p_{\text{ref}}$ and (3) evaluating the remaining $1 - (\delta_1 + \delta_2)$ tuples by $p^{\star}$.
Complete algorithm for each phase is shown in  Appendix~\ref{sec:appendix:algorithm}. In the following, we elaborate on the key design of each phase.

\begin{algorithm}[t]
\small
\caption{{\small Adaptive Query Execution}}
\label{alg:aqe-main}
\SetKwInOut{Input}{Input} \SetKwInOut{Output}{Output}
\Input{Input chunks $\mathcal{D}$, expr\_limit $\delta_1 \cdot |\mathcal{D}|$ , path\_limit  $\delta_2 \cdot |\mathcal{D}|$}
 $n \leftarrow 0$ \;
\For{chunk $c \in \mathcal{D}$}{ 
    \If{$n < \text{expr\_limit}$}{ \label{line:phase1:start}
        collect statistics of filters of $c$ \algocomment{Phase 1: Exp. Exploration} \label{line:phase1:end}\; 
    }
    \ElseIf{$n < \text{path\_limit}$}{ \label{line:phase2:start}
        \If{paths not generated}{
        generate candidate paths $\mathcal{P}$ using the statistics of $c$ \algocomment{Phase 2: Path Exploration} \;
        }
        evaluate paths in $\mathcal{P}$ on $c$ and collects their metics  \;
    }
  \Else{
    \If{optimal path not selected}{
        select optimal path $p^\star$ based on collected metrics  \label{line:phase2:end} \;
    }
    execute optimal path $p^\star$ on $c$ \algocomment{Phase 3: Path Exploitation} \label{line:phase3} \;
    }
    $n \leftarrow n + |c|$ \;
} 
\end{algorithm}

\subsubsection*{Phase 1: Expression Exploration} 
Given a sequence of \SemFilter $\Sigma = \{\sigma_1, \cdots, \sigma_n\}$, in the expression exploration phase, we execute each \SemFilter $\sigma_i$ independently on a fraction of $\delta_1$ tuples, obtaining its selectivity $s_i$ and query result $R_i$, denoted as a boolean vector.
The query results are used to evaluate the potential of operator fusion. Specifically, the higher the correlations between two filters, the more likely the fusion can preserve the semantics.
We use the Matthews Correlation Coefficient (MCC)~\cite{MATTHEWS1975442} to quantify the pair-wise correlation of two \SemFilter, statistically from the perspective of their results, where $\text{MCC}(\sigma_i, \sigma_j)$ equals to $1$ and $-1$ indicates an absolute positive and negative correlation between $\sigma_i$ and $\sigma_i$, respectively, and $\text{MCC}(\sigma_i, \sigma_j)$ equals to $0$ indicates a random association. Only a pair of \SemFilter whose MCC is larger than a threshold is considered to be fused into one \SemFilter.

\subsubsection*{Phase 2: Path Exploration}
Based on the statistics collected in Phase 1, we first generate a set of candidate paths $\mathcal{P}$, including the reference path $p_{\text{ref}}$ and a base path ${p}_{\text{base}}$. 
The reference path $p_{\text{ref}}$ preserves the order of $\Sigma$ as the user-specified order in the \SemaSQL query, while the base path ${p}_{\text{base}}$ reorders the filters in $\Sigma$ ascendingly by their selectivities.
For operator fusion, if $\text{MCC}(\sigma_i, \sigma_j)$ is larger than a threshold $\tau$, we merge the two \SemFilter operators $\sigma_i, \sigma_j$ into one \SemFilter $\sigma_{i \oplus j}$, and treat its selectivity $s_{i\oplus j}$ as the minimum of $s_i$ and $s_j$.
For $n$ \SemFilter operators, we only consider introducing one fusion into candidate paths, to prevent degeneration of query accuracy and explosion of exploration space, resulting in at most $n(n - 1)/2$ candidate paths with $n - 1$ operators ordering by selectivity ascendingly.
For all paths with fused operators, as well as the reference path and base path, we also generate their counterpart paths with batch prompting.

Next, the executor evaluates these candidate paths on a fraction of $\delta_2$ tuples, collecting the query latency, token consumption, and the query results.  
As the ground-truth of semantic queries is unavailable beforehand, we treat the query results of ${p}_{\text{ref}}$ without batch prompting as proxy ground-truth, which is used to evaluate the accuracy of other candidate paths. 
If the accuracy, e.g., F1 or Accuracy, is below a threshold $\tau_{acc}$, it indicates that the corresponding path generates results far divergent from the reference path, and that path will be discarded to guarantee high result quality. 
During path evaluation, we devise two dynamic pruning strategies to reduce the exploration cost. 
First, if the accuracy of the reference path with batch prompting is below the threshold, the system will not explore all paths with batch prompting. 
Second, if the accuracy of one candidate path with operator fusion cannot meet the threshold, the evaluation of its counterpart with batch prompting will be bypassed.
After the candidate paths are evaluated, we compute the Pareto frontier in the bi-objective space of query latency and token cost. Regarding user or application preference, the latency-first or token-first optimal path is selected as the optimal path $p^\star$.

\subsubsection*{Phase 3: Path Exploitation}
The selected path $p^\star$ is applied to process the remaining bulk of the data. Since the plan has been validated for both efficiency and accuracy, this phase ensures predictable, optimized performance while amortizing the cost of path exploration.

We present a case study of AQE in~\cref{sec:exp:breakdown_analysis} on a specific query with three \SemFilter operators, serving as an illustrative example. 
Currently, AQE is only supported for \SemFilter operators in one execution pipeline of DuckDB. Optimizing more complex operators at runtime, especially for those across multiple pipelines will incur over-sophisticated implementation and large overhead of context switch, which deserves deeper engineering optimization.

\section{Engineering Details}
\label{sec:engineering_details}
We present key engineering details of \Sema, focusing on concurrent LLM calls and pipeline-parallel adaptive query execution (AQE).

\stitle{Concurrent LLM Invocations.}
To integrate LLM inference into the system, \Sema extends DuckDB's pipeline-parallel engine  with a thread-local concurrent I/O design.
DuckDB uses a morsel-driven execution framework~\cite{DBLP:conf/sigmod/LeisBK014}, decomposing work into fine-grained morsels, each processing a small batch of tuples within a pipeline. 
A central scheduler dynamically assigns morsels to worker threads from a shared pool, ensuring balanced load distribution and efficient CPU utilization with minimal synchronization overhead.
In \Sema, each worker thread maintains a dedicated CURL multi-handle instance as its private HTTP connection pool. During execution, the thread synchronously issues LLM API requests within the morsel loop, while the CURL multi-handle enables non-blocking progress and efficient connection reuse. This design avoids cross-thread synchronization, allowing multiple workers to issue and manage LLM requests concurrently.
As a result, \Sema preserves DuckDB’s fine-grained parallelism and fully utilizes its task scheduler for both computation and I/O-bound workloads.

\stitle{Pipeline-Parallel AQE.}
\Sema integrates AQE directly into DuckDB’s pipeline-parallel engine via a customized operator, \textit {PhysicalSemAdaptiveFilter}, which unifies adaptive decision-making with streaming execution. Each pipeline executor incrementally pulls data chunks and processes them through the adaptive three-phase 
execution process of \Sema, dynamically selecting and applying the optimal execution path. Leveraging DuckDB’s streaming primitives (\texttt{NEED\_MORE\_INPUT} and  \texttt{HAVE\_MORE\_OUTPUT}), \Sema performs progressive data slicing, evaluating only small fractions of the input during exploration before committing to the optimal path for the remaining data.

During execution, multiple concurrent executors operate in parallel  on disjoint data partitions, each maintaining thread-local states (including expression evaluators and intermediate buffers) while reporting lightweight performance metrics (e.g., selectivity, latency, and token-level cost) to a shared global state. This dual-state architecture enables global coordination with local autonomy: adaptive decisions are made collectively yet executed independently, minimizing synchronization overhead. 
As pipelines progress, the global controller aggregates feedback via atomic updates and propagates refined execution choices to all threads. Consequently, \Sema achieves fine-grained adaptivity and stable throughput across varying data distributions, remaining compatible with the lightweight, streaming, and memory-efficient execution model of DuckDB.

\section{System Evaluation}
\label{sec:exp}
We present the experimental setup in \cref{sec:exp:setup} and report our comprehensive evaluation in the following facets:
\ding{172} Compare \Sema with baseline systems on output quality, efficiency (\cref{sec:exp:overall});
\ding{173} Assess the impact of operator fusion and prompt batching in \Sema (\cref{sec:exp:execution_optimization});
\ding{174} Explore the utility of LLM-assisted expression optimizations (\cref{sec:exp:rule-based optimization}); 
\ding{175} Study the effectiveness of breakdown optimizations (\cref{sec:exp:breakdown_analysis}).

\subsection{Experimental Setup}
\label{sec:exp:setup}
\stitle{Datasets.}
We use 9 datasets from BIRD~\cite{DBLP:conf/nips/BIRD-Benchmark}, a widely used benchmark originally designed for NL2SQL. Table~\ref{tab:datasets} in Appendix~\ref{sec:appendix:benchmark} summarizes the profiles of the datasets, where each dataset represents distinct analytical applications. 
Specifically, \textit{Appstore}, \textit{Superstore}, and \textit{Food\_Inspection} focus on entity-level semantic extraction, while others like \textit{Movies\_4} and \textit{Authors} emphasize multi-table reasoning and text-intensive query evaluation. 
This diversity ensures robust validation of \Sema's optimization strategies across both structured and unstructured data.

\stitle{Benchmark Queries.} Based on the 9 datasets, we design 20 semantic queries, forming a comprehensive testbed for semantic query processing. 
The queries fall into three categories: 
\ding{182} Single-operator queries (Q1-Q5): contain a single \SemFilter or \SemProj.
\ding{183} Consecutive \SemFilter queries (Q6-Q10): contain 2-3 consecutive \SemFilter operators, used to evaluate the AQE framework and execution optimizations.
\ding{184} Queries with consecutive semantic operators other than \SemFilter (Q14-Q20): include queries with \SemProj and \SemFilter (Q14-Q15), queries with two \SemProj operators (Q14-Q15), queries with \SemFilter and \SemProj (Q16-Q18), and queries with \SemFilter and \SemAgg (Q19-Q20). The complete 20 queries in \SemaSQL are presented in Appendix~\ref{sec:appendix:benchmark}, Fig.~\ref{fig:q1}-\ref{fig:q20}.
Ground-truth results for Q1-Q5 are collected from the original datasets, while those for  Q6-Q20 are obtained from the inference results of a reference LLM (Gemma-3 27B by default).

\stitle{Evaluation Metrics.}
For quality, we employ a range of evaluation metrics regarding the specific objectives of each query:
\ding{182} Q1-Q14: Single label, multi-class classification; we report Accuracy, Precision, Recall, and F1.
\ding{183} Q15: Multi-label, multi-class classification. We report subset Accuracy, micro- and macro- F1 and Jaccard, respectively.
\ding{184} Q16-Q20: Each involves two consecutive operators. We first evaluate the binary classification performance of the \SemFilter, then assess the downstream task. Specifically, Q16 is a keyword extraction task (evaluated by macro- Precision/Recall/F1 and Jaccard); Q17 and Q18 are classification tasks (weighted-F1, micro-F1, macro-F1); Q19 and Q20 focus on aggregation and summarization (measured by word overlap between generated and ground-truth results).
For time and cost efficiency, we use query latency, \# LLM invocations, \# input/output tokens, and cost (\textcent) of API calls as evaluation metrics.

\stitle{Baselines.}
We compare \Sema against three representative systems: \Lotus~\cite{patel2024lotus} built on Pandas, with optimized algorithms for \SemFilter and \SemJoin, \Palimpzest~\cite{liu2025palimpzest} with a cascaded optimizer and \FlockMTL~\cite{DBLP:journals/corr/abs-2504-01157}, which implements semantic operators as UDFs within DuckDB. The capabilities of these baseline systems and \Sema regarding vectorized execution, pipeline parallelism and asynchronous execution are shown as below:

\begin{table}[h]
\begin{center}
\label{tab:aqe_path_statistics}
\footnotesize
\begin{tabular}{c lccc}
\toprule
Capability & \Lotus & \FlockMTL & \Palimpzest  & \Sema \\
\midrule
Vectorized Execution & \cmark & \xmark & \xmark & \cmark \\
Pipeline Parallelism &  \xmark & \cmark  & \cmark & \cmark \\
Asynchronous Execution & \cmark & \xmark & \cmark & \cmark \\
\bottomrule
\end{tabular}
\end{center}
\end{table}

\stitle{Implementation Details.}
\Sema is built on DuckDB v1.2.2, supporting both local and remote LLM invocations.  We use vLLM~\cite{kwon2023efficientmemorymanagementlarge} as the local LLM inference engine and \OpenRouter as remote LLM APIs, respectively. We deploy Google’s Gemma-3 model \cite{gemmateam2025gemma3technicalreport} with 4B, 12B, and 27B, with a temperature of 0 for inference reproducibility.
LLMs with a temperature of 0 will generate the same query results regardless of the reordering of \SemFilter.
For vLLM, all models are configured with \kw{tensor\_parallel\_size}=4, bfloat16 precision, and a GPU memory utilization limit of 0.9. 
Local experiments are conducted on a Linux server with four NVIDIA RTX 5090 GPUs and an AMD EPYC 9554 CPU (64 cores, 503 GB RAM).
Pricing for remote API calls follows DeepInfra's rate on \OpenRouter: $0.04\$$/million input tokens and $0.13\$$/million output tokens for Gemma-3 12B, $2\$$/million input tokens and $8\$$/million output tokens for GPT-4.1.

\subsection{Overall Evaluation of Quality \& Efficiency}

We compare the output quality and query latency of \Sema against three baseline systems using 20 semantic queries. To reduce the influence of network delay and fluctuation, local vLLM is used for all the systems. For Q1–Q5, we use the Gemma-3 4B model since these queries involve only a single operator and are relatively simple, while for the more complex queries Q6–Q20 that consist of multiple operators, we adopt the Gemma-3 12B model. For \Palimpzest, we use \MinTime optimizer policy to execute all queries.

\label{sec:exp:overall}

\begin{figure*}
    \centering
    \includegraphics[width=0.95\linewidth]{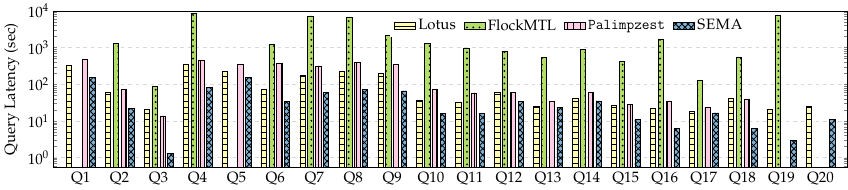}
    \vspace{-2ex}
    \caption{Overall Query Latency (seconds)}
    \label{fig:latency_all}
\end{figure*}

\begin{table*}[t]
    \centering
    \setlength{\tabcolsep}{1.8pt}
    \tiny
    \caption{Query Accuracy for Q1-Q14 (\FlockMTL does not support Q1, Q5 and Q20.)}
    \vspace{-2ex}
    \label{table:acc_q1_q14}
    \begin{tabular}{c|cccc|cccc|cccc|cccc|cccc|cccc|cccc}
    \toprule
        \multirow{2}{*}{\textbf{System}} & \multicolumn{4}{c|}{\textbf{Q1}} & \multicolumn{4}{c|}{\textbf{Q2}} & \multicolumn{4}{c|}{\textbf{Q3}} & \multicolumn{4}{c|}{\textbf{Q4}} & \multicolumn{4}{c|}{\textbf{Q5}} & \multicolumn{4}{c|}{\textbf{Q6}} & \multicolumn{4}{c}{\textbf{Q7}} \\
        & Acc. & Pre. & Rec. & F1 & Acc. & Pre. & Rec. & F1 & Acc. & Pre. & Rec. & F1 & Acc. & Pre. & Rec. & F1 & Acc. & Pre. & Rec. & F1 & Acc. & Pre. & Rec. & F1 & Acc. & Pre. & Rec. & F1 \\ \midrule
        Lotus & 0.9940 & 1.0000 & 0.9940 & 0.9970 & 0.7367 & 0.8148 & 0.7367 & 0.7450 & 0.7157 & 1.0000 & 0.7157 & 0.8343 & 0.4194 & 0.8837 & 0.4194 & 0.4534 & 0.3040 & 1.0000 & 0.3040 & 0.4662 & 0.6977 & 1.0000 & 0.2561 & 0.4078 & 0.8489 & 0.8722 & 0.6102 & 0.7180 \\
        FlockMTL & - & - & - & - & 0.7973 & 0.8221 & 0.7973 & 0.7994 & 0.2294 & 1.0000 & 0.2294 & 0.3732 & 0.4897 & 0.7182 & 0.4897 & 0.4543 & - & - & - & - & 0.7195 & 0.9023 & 0.3475 & 0.5017 & 0.7504 & 0.9558 & 0.2185 & 0.3557 \\
        \Palimpzest & 0.9461 & 1.0000 & 0.9461 & 0.9723 & 0.3832 & 0.4471 & 0.3832 & 0.4022 & 0.9426 & 1.0000 & 0.9426 & 
        0.9705 & 0.2573 & 0.4080 & 0.2573 & 0.2326 & 0.5799 & 1.000 & 0.5799 & 0.7341 & 0.7011 & 0.9928 & 0.2664 & 0.4201 & 0.8556 & 0.8316 & 0.6796 & 0.7480 \\
        \Sema & 0.9934 & 1.0000 & 0.9934 & 0.9967 & 0.7369 & 0.8154 & 0.7369 & 0.7407 & 0.7124 & 1.0000 & 0.7124 & 0.8320 & 0.4428 & 0.7349 & 0.4428 & 0.4878 & 0.1265 & 1.0000 & 0.1265 & 0.2245 & 0.7030 & 1.0000 & 0.2693 & 0.4243 & 0.8148 & 0.7825 & 0.5715 & 0.6606 \\
    \midrule
        \multirow{2}{*}{\textbf{System}} & \multicolumn{4}{c|}{\textbf{Q8}} & \multicolumn{4}{c|}{\textbf{Q9}} & \multicolumn{4}{c|}{\textbf{Q10}} & \multicolumn{4}{c|}{\textbf{Q11}} & \multicolumn{4}{c|}{\textbf{Q12}} & \multicolumn{4}{c|}{\textbf{Q13}} & \multicolumn{4}{c}{\textbf{Q14}} \\
        & Acc. & Pre. & Rec. & F1 & Acc. & Pre. & Rec. & F1 & Acc. & Pre. & Rec. & F1 & Acc. & Pre. & Rec. & F1 & Acc. & Pre. & Rec. & F1 & Acc. & Pre. & Rec. & F1 & Acc. & Pre. & Rec. & F1 \\ \midrule
        Lotus & 0.8542 & 0.6728 & 0.8921 & 0.7670 & 0.9959 & 0.5000 & 0.1154 & 0.1875 & 0.9456 & 0.9412 & 0.3743 & 0.5356 & 0.8697 & 0.9887 & 0.3954 & 0.5648 & 0.7358 & 0.6534 & 0.4096 & 0.5035 & 0.7181 & 0.8864 & 0.6754 & 0.7666 & 0.7168 & 0.7767 & 0.7168 & 0.7115 \\
        FlockMTL & 0.7445 & 0.9706 & 0.0574 & 0.1083 & 0.9962 & 1.0000 & 0.0769 & 0.1429 & 0.9251 & 0.6809 & 0.1882 & 0.2949 & 0.1563 & 0.0870 & 0.3103 & 0.1360 & 0.6613 & 0.4722 & 0.2993 & 0.3664 & 0.7440 & 0.8196 & 0.8035 & 0.8115 & 0.7059 & 0.7684 & 0.7059 & 0.7249 \\
        \Palimpzest & 0.8644 & 0.7979 & 0.6680 & 0.7272 & 0.9956 & 0.4167 & 0.1923 & 0.2632 & 0.9407 & 0.5953 & 0.9000 & 0.7166 & 0.9381 & 0.8503 & 0.8621 & 0.8562 & 0.6855 & 0.5168 & 0.5951 & 0.5532 & 0.7305 & 0.8441 & 0.7445 & 0.7912 & 0.6517 & 0.6763 & 0.6517& 0.6543 \\
        \Sema & 0.8426 & 0.6958 & 0.7375 & 0.7160 & 0.9955 & 0.3636 & 0.1538 & 0.2162 & 0.9543 & 0.9364 & 0.4873 & 0.6410 & 0.9651 & 0.9740 & 0.8598 & 0.9133 & 0.7273 & 0.6082 & 0.4683 & 0.5292 & 0.7315 & 0.9012 & 0.6834 & 0.8883 & 0.7570 & 0.7954 & 0.7570 & 0.7563 \\
    \bottomrule
    \end{tabular}
\end{table*}

\begin{figure*}
    \begin{tabular}[h]{c}
        \subfigure[Q15] {
				\includegraphics[width=0.32\columnwidth]{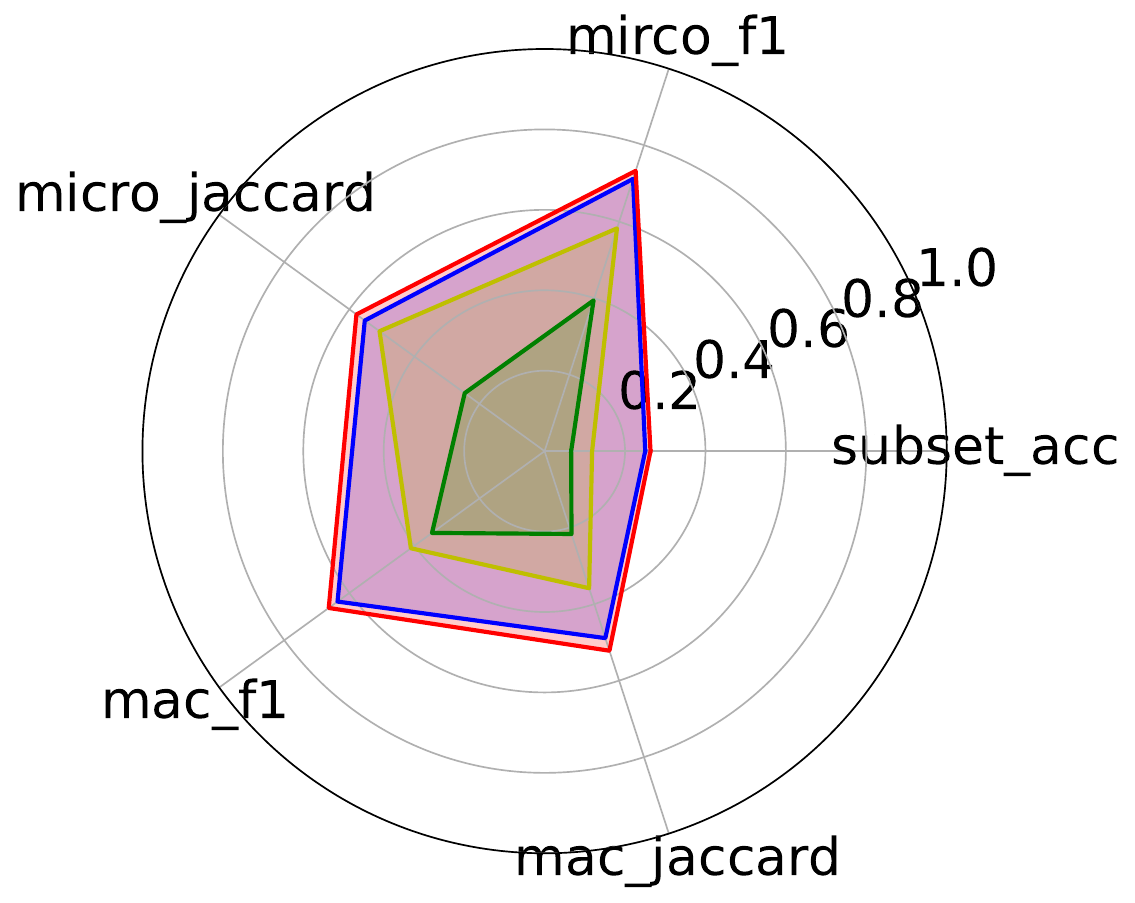}
			\label{fig:T_heatmap}
		} 
        \hspace{-1ex}
        \subfigure[Q16] {
			\includegraphics[width=0.32\columnwidth]{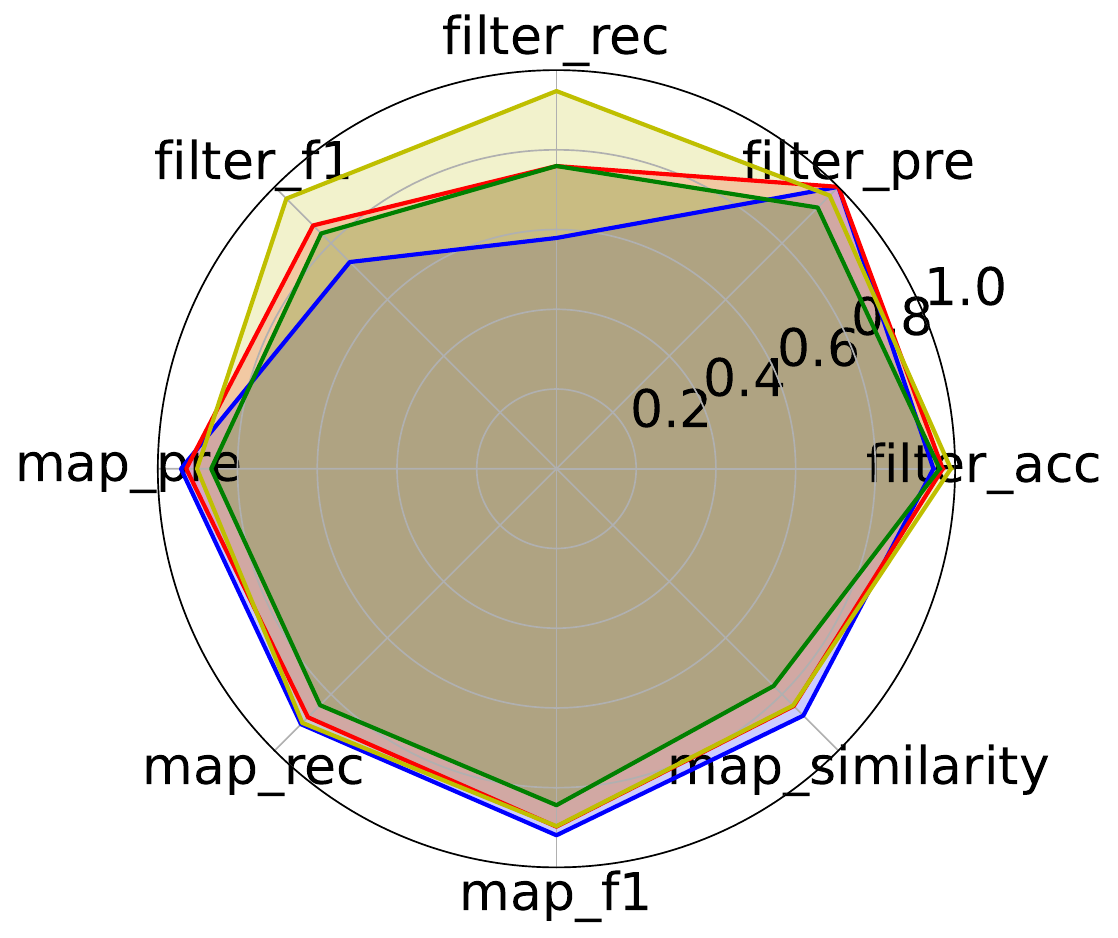}
			\label{fig:F_heatmap}
		}
        \hspace{-1ex}
        \subfigure[Q17] {
				\includegraphics[width=0.335\columnwidth]{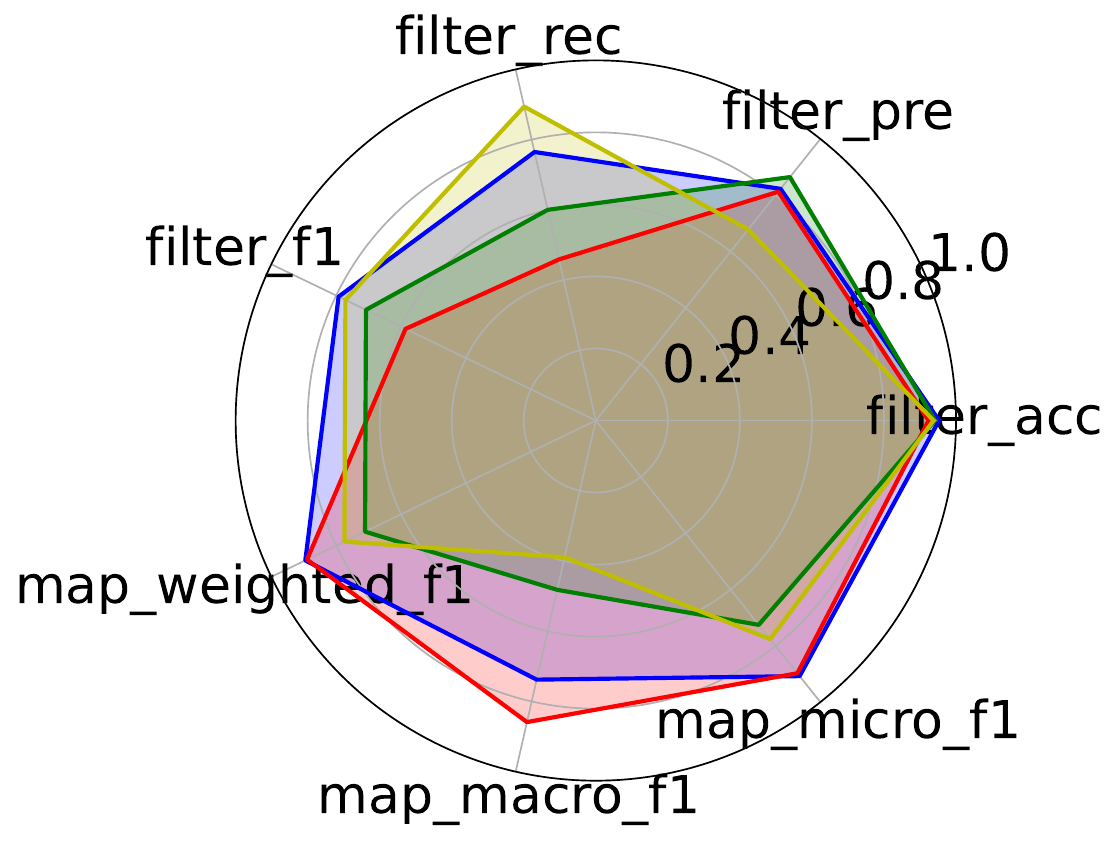}
			\label{fig:T_heatmap}
		} 
        \hspace{-1ex}
        \subfigure[Q18] {
			\includegraphics[width=0.335\columnwidth]{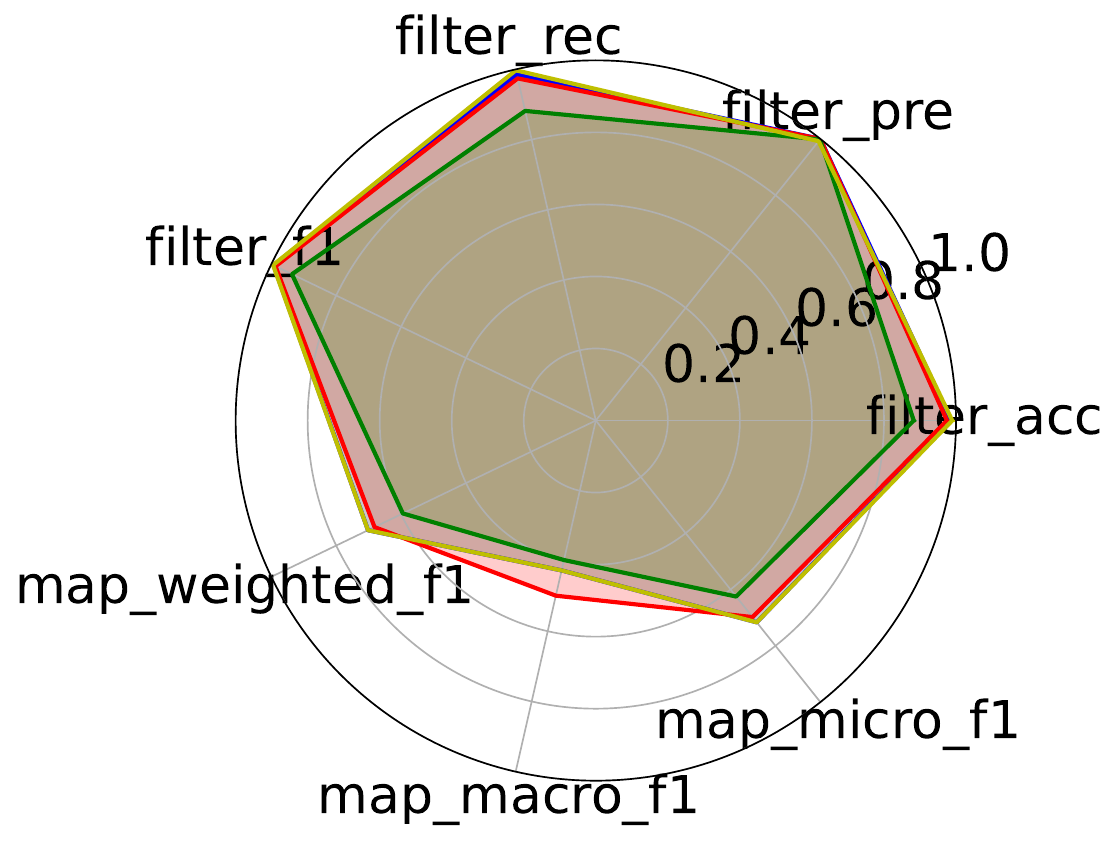}
			\label{fig:F_heatmap}
		}
        \hspace{-1ex}
        \subfigure[Q19] {
            \includegraphics[width=0.32\columnwidth]{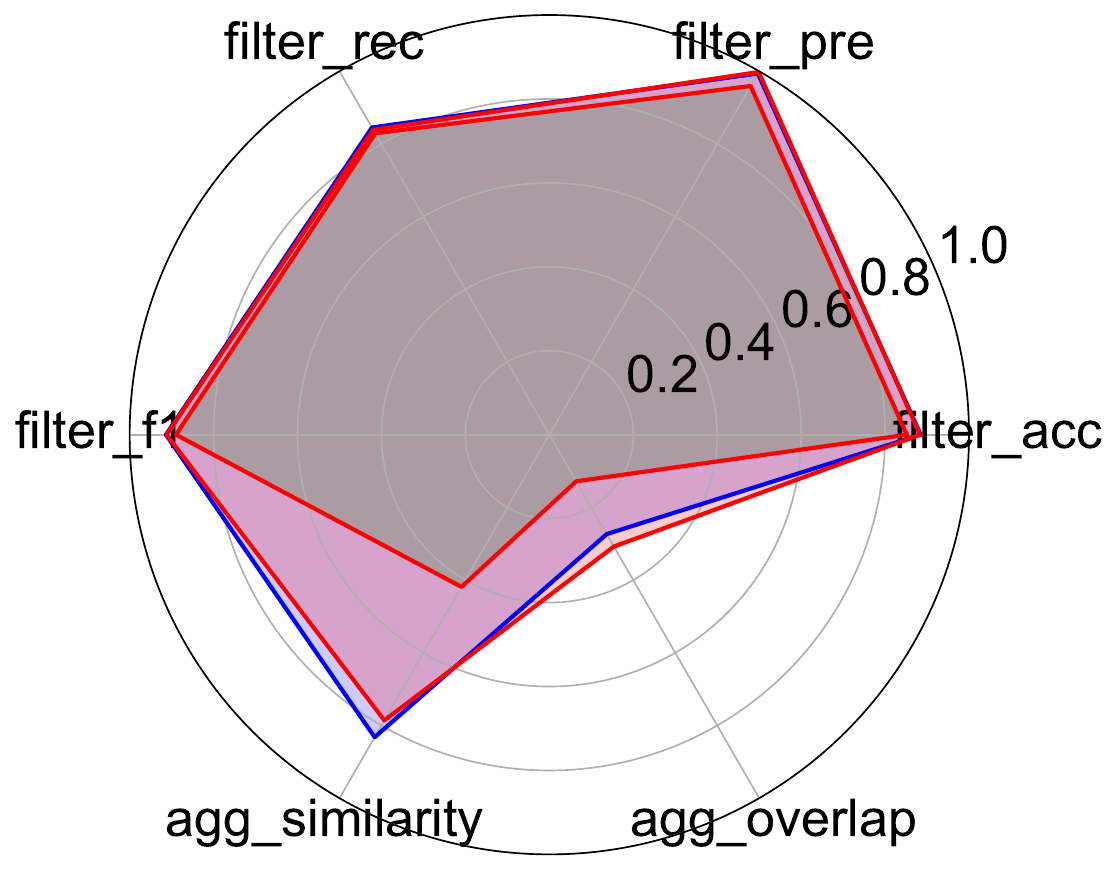}
            \label{fig:T_heatmap}
        }
        \hspace{-1ex}
        \subfigure[Q20] {
			\includegraphics[width=0.32\columnwidth]{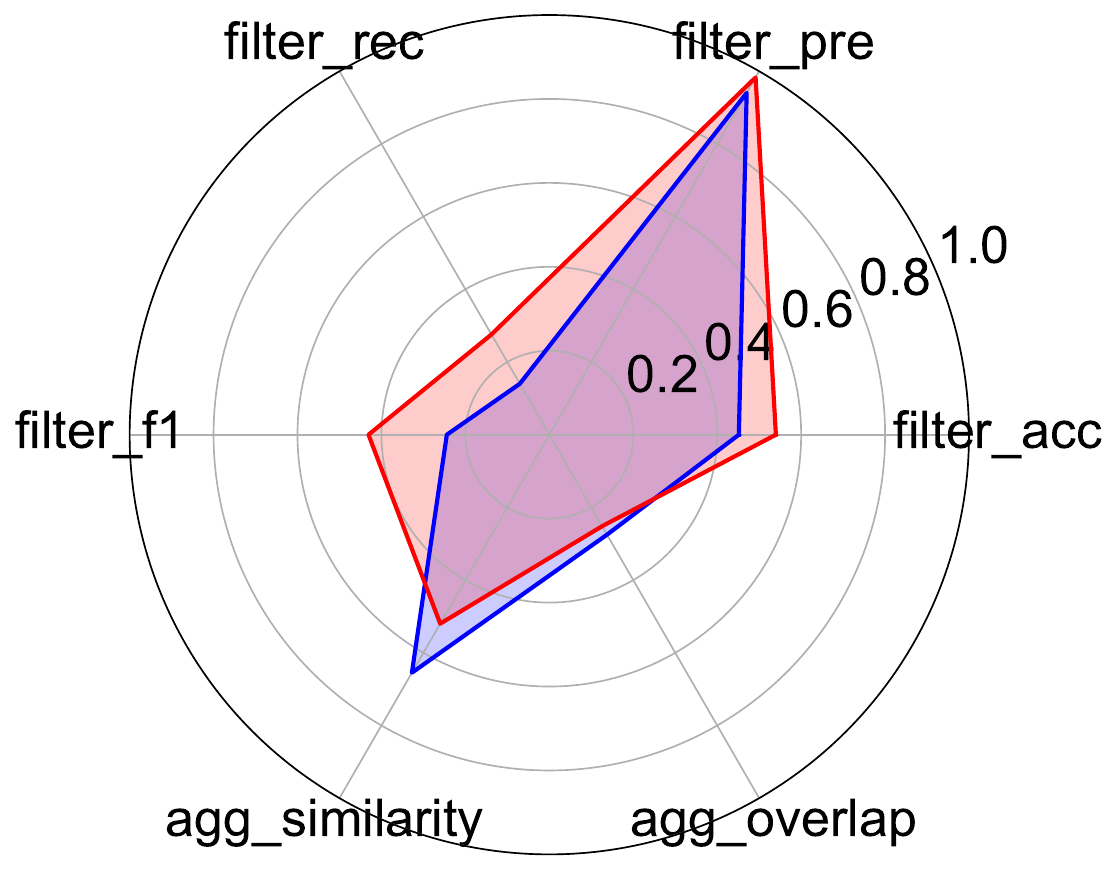}
			\label{fig:F_heatmap}
		}
	\end{tabular}
    \vspace{-2ex}
	\caption{Query Accuracy of {\color{blue} Lotus}, {\color{red} \Sema}, {\color{yellow} \Palimpzest} and {\color{Green} FlockMTL} for Q15-Q20}
        \label{fig:Q15-Q20 quality}
\end{figure*}

\stitle{Overall Efficiency Comparison.}
As shown in Fig.~\ref{fig:latency_all}, \Sema consistently outperforms all three baseline systems in query latency across all 20 queries. 
Particularly, the performance advantage of \Sema is remarkable for complex  queries (Q6–Q20), where \Sema achieves up to a 3–6$\times$ speedup. This improvement is attributed to \Sema's vectorized execution, pipeline parallelism and asynchronous execution mechanisms, which allow it to  effectively utilize hardware concurrency and effectively interplay with vLLM during LLM invocation. In contrast, the three baseline systems exhibit notable limitations stemming from missing execution optimizations. \Lotus, although supporting vectorized and asynchronous execution, lacks pipeline parallelism, which prevents effective overlap between operators. \FlockMTL suffers from more fundamental inefficiencies: it neither leverages vectorized execution nor supports asynchronous execution, leading to inefficient resource usage and slower execution. \Palimpzest, while enabling pipeline parallelism and asynchronous execution, does not exploit vectorized execution, limiting its ability to benefit from a high-throughput execution model and consequently constraining overall performance. \FlockMTL fails to support queries Q1, Q5, Q19, and Q20 due to the presence of columns containing multiple data types. \Palimpzest is unable to execute query Q20, as it currently lacks support for semantic aggregation with \SemAgg.

\stitle{Overall Quality Comparison.}
Table~\ref{table:acc_q1_q14} and Fig.~\ref{fig:Q15-Q20 quality} show that \Sema, \Lotus and \Palimpzest achieve comparable accuracy across most evaluation metrics, with each system demonstrating strengths on different queries. 
This variation is largely due to the prompt sensitivity of LLM outputs, as we observed in the testing that minor differences in prompt handling can lead to noticeable fluctuations in the results.
Although the prompt format of \Sema is carefully aligned with \Lotus to ensure a fair comparison, slight discrepancies remain due to differences in system architecture and execution flow. 
In general, \Sema maintains balanced and competitive performance with \Lotus and \Palimpzest on both simple and complex queries. In contrast, \FlockMTL performs considerably worse across almost all metrics, reflecting the limitations in its model invocation strategy and overall system robustness.

\newcolumntype{L}{>{\centering\arraybackslash}p{2.3cm}} 
\newcolumntype{T}{>{\raggedright\arraybackslash}p{2cm}}

\subsection{Execution Optimization Evaluation}
\label{sec:exp:execution_optimization}

To assess the utility of execution optimizations, we compare the performance of \Sema under four configurations: (1) \textit{w/o  optimization}, (2) \textit{w/ prompt batching} with a batch size of 16, (3) \textit{w/ operator fusion}, and (4) \textit{w/ batch \& fusion}. All experiments are conducted using Gemma-3 12B, invoked both remote API and locally. We use four queries, Q8, Q13, Q14 and Q16 with distinct combinations of operators to evaluate the above configurations.

\begin{figure}
    \begin{tabular}[t]{c}
        \centering
        \includegraphics[width=0.9\columnwidth]{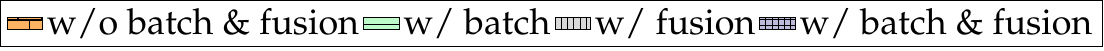}
	\end{tabular}
    \begin{tabular}[h]{c}
    \subfigure[Latency (local)]{
        
         \includegraphics[width=0.45\columnwidth]{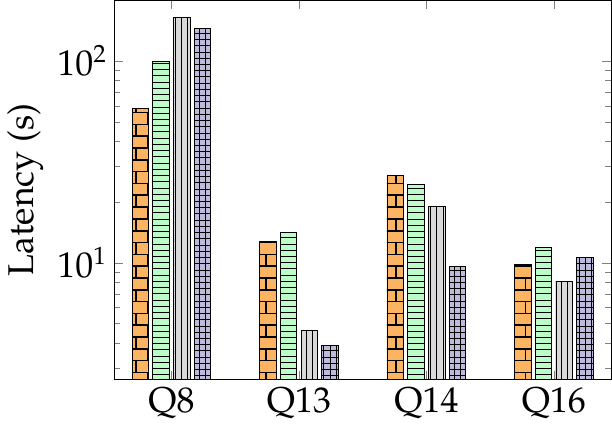}
         \label{fig:exe:latency:local}
    }
    
    \subfigure[Latency (remote)]{
        \includegraphics[width=0.45\columnwidth]{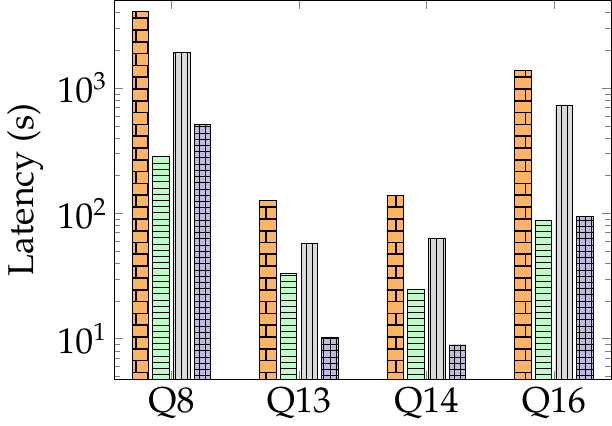}
        \label{fig:exe:latency:remote}
        
    }
    \\
    \subfigure[Cost (local)]{
        \includegraphics[width=0.45\columnwidth]{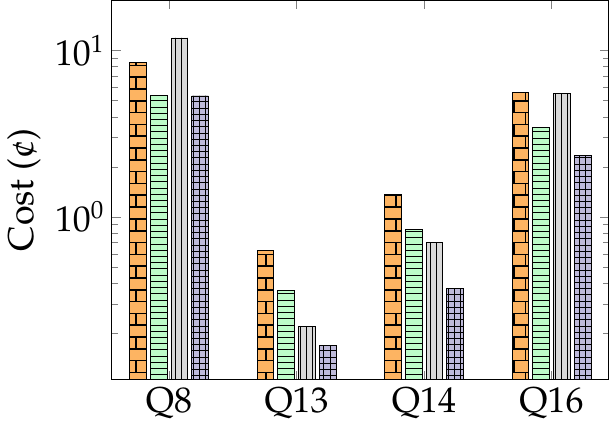}
        \label{fig:exe:cost:remote}
    }
    \subfigure[Cost (remote)]{
        \includegraphics[width=0.45\columnwidth]{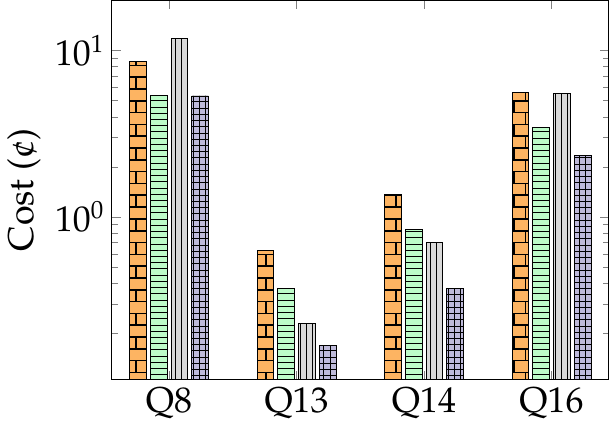}
        \label{fig:exe:cost:remote}
    }
    \\
    \subfigure[F1 (local)]{
        \includegraphics[width=0.45\columnwidth]{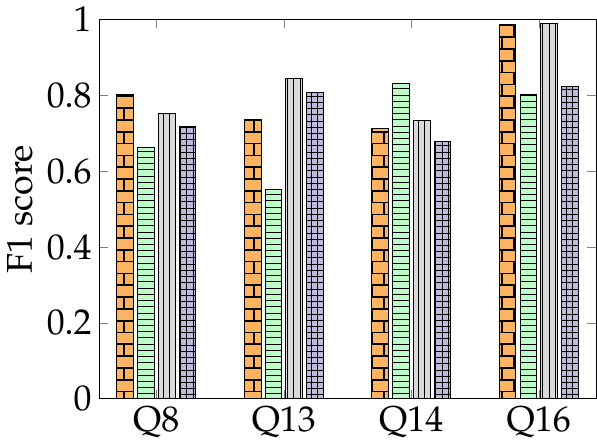}
        \label{fig:exe:f1:local}
    }
    \subfigure[F1 (remote)]{
        \includegraphics[width=0.45\columnwidth]{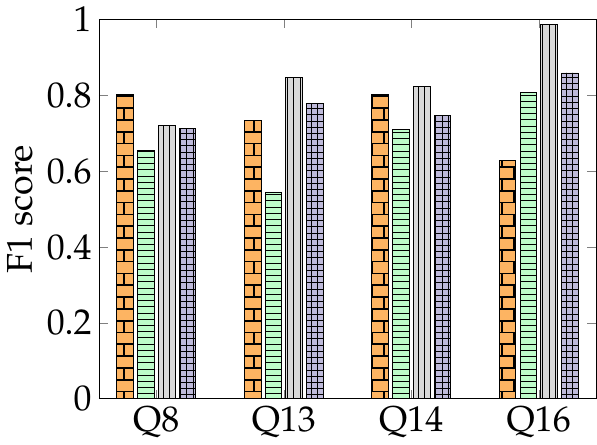}
        \label{fig:exe:f1:remote}
    }
    \end{tabular}
    \vspace{-3mm}
    \caption{The Effectiveness of Execution Optimizations}
    \label{fig:execution optimization}
\end{figure}

The experimental results are summarized in Fig.\ref{fig:execution optimization}, which compares the latency, execution cost and output quality of prompt batching and operator fusion across different workloads under both remote and local deployment settings. In terms of latency, both batching and fusion reduce overall execution time in the remote setting because they decrease the number of LLM invocations and therefore amortize network and scheduling overhead. Since batching aggregates more requests into a single call, it consistently achieves lower latency than fusion. However, in the local setting, batching and fusion may increase latency. Both techniques trade fewer invocations for longer input and output sequences, while the vLLM continuous batching scheduler is optimized for many short requests rather than fewer long ones, leading to reduced scheduling efficiency.
Regarding cost, prompt batching and operator fusion generally lower query execution cost by reducing redundant input tokens processed by the model. An exception appears in Q8, where fusion fails to reduce cost and may even increase it. This is mainly because fusion combines operations involving a filter operator; when the filter predicate's selectivity is low, additional unnecessary tokens must still be processed, offsetting potential savings.
For output quality, batching consistently degrades performance due to interference among multiple queries handled simultaneously within a single prompt. Fusion shows task-dependent behavior: it improves quality for Q13, Q14, and Q16 but degrades it for Q8. The reason is that Q8 consists of independent tasks, where joint processing introduces interference, whereas Q13, Q14, and Q16 are sequentially dependent tasks whose shared context allows the model to reuse intermediate reasoning and achieve better results.

\subsection{Expression Optimization Evaluation}
\label{sec:exp:rule-based optimization}
We conduct effectiveness studies of the two NL expression optimizations, expression compression and predicate deduction, on Q3, Q6, Q13, and Q14, and the results are summarized in Table~\ref{table:exp_opt}. To validate expression compression, we intentionally introduced semantically redundant content into the NL expressions of the queries. To validate partial deduction, we rewrote certain SQL predicates as NL expressions to assess whether predicate deduction can effectively achieve deduction. These query variants can be found in Appendix~\ref{sec:appendix:benchmark}, Fig.~\ref{fig:q3:beforeer}-\ref{fig:q14-beforepd}.

\textbf{Expression compression (EC)} primarily targets token efficiency by reducing redundancy in NL expressions. Since, EC rewrites verbose NL expressions into more compact yet semantically equivalent forms, enabling LLM to operate on shorter prompts, both input and output token consumption are reduced while maintaining comparable execution quality. A representative example can be observed in Q3, where the input tokens decrease from 73.24K (w/o EC) to 72.20K and the output tokens decrease from 1.07K to 1.00K after applying EC, while precision, recall, and F1 remain nearly unchanged. Similar patterns can be observed in Q6 and Q13, suggesting that EC effectively improves token efficiency without introducing semantic distortion or degrading result quality.

\textbf{Predicate deduction (PD)} focuses on decreasing the number of required LLM calls during query execution. PD derives necessary SQL predicates directly from the NL expressions and injects them into the execution process, thereby filtering out tuples before LLM invocation. The benefit is particularly evident in Q13, where introducing PD brings efficiency with fewer calls and reduced overhead while maintaining comparable result quality.

\begin{table}[!t]
\centering
\scriptsize
\setlength{\tabcolsep}{4pt} 
\caption{Impact of Expression Optimization. Here, `+' adds the invocation costs in {\color{red} red} of the auxiliary LLM)}
\label{table:exp_opt}
\begin{tabular}{lrccc r ccc}
\toprule
\textbf{Q} & \textbf{Opt.} & \textbf{Pre.} & \textbf{Rec.} & \textbf{F1}  &\textbf{Lat. (s)}& \textbf{IT (K)}& \textbf{OT (K)} & \textbf{LLM Calls} \\
\midrule

\multirow{4}{*}{Q3}
 & w/o EC & 1.0000 & 0.9880 &  0.9944 &95.43
&  91.24& 1.08& 401
\\
 & w/ EC & 1.0000 & 0.9890 & 0.9945 &102.05
& 72.20+\textcolor{red}{1.69}& 1.00+\textcolor{red}{0.09}& 401+\textcolor{red}{1}
\\
 & w/o PD & 1.0000 & 0.9950 &  0.9975 & 410.66
&  191.24& 6.08 & 856
\\
 & w/ PD & 1.0000 & 0.9950 & 0.9975 & 135.11
& 73.24+\textcolor{red}{1.07}& 1.07+\textcolor{red}{0.051}& 492+\textcolor{red}{2}
\\
\midrule

\multirow{4}{*}{Q6}
 & w/o EC & 0.9868
& 0.9418
& 0.9637
&256.42
& 9.16& 0.15& 59
\\
 & w/ EC & 0.9818
& 0.9795
& 0.9803
&239.32& 8.15+\textcolor{red}{1.46}& 0.14+\textcolor{red}{0.03}& 59+\textcolor{red}{1}
\\
& w/o PD & 1.0000
& 0.9072
& 0.9513
&443.66
& 7.76& 0.16& 59
\\
& w/ PD & 1.0000
& 0.9072
& 0.9513
&447.67
& 7.76 + \textcolor{red}{1.13} & 0.16 + \textcolor{red}{0.03} & 59+\textcolor{red}{2}
\\

\midrule

\multirow{4}{*}{Q13}
 & 
w/o EC & 0.8112 & 0.4610& 0.5878&342.25& 38.73& 3.98& 318\\
 & 
w/ EC & 
0.7647& 0.6569& 0.7067&357.97& 34.84+\textcolor{red}{3.38}& 4.28+\textcolor{red}{0.07}& 318+\textcolor{red}{1}\\
 & w/o PD      & 0.8057 & 0.5293& 0.6388& 637.30& 57.10& 4.62& 601\\
 & w/ PD  & 0.8182 & 0.5505& 0.6582& 346.65 & 31.64+\textcolor{red}{1.25}& 3.90+\textcolor{red}{0.03}& 318+\textcolor{red}{2}\\
 \midrule
 
 \multirow{4}{*}{Q14}
 & w/o EC & 0.5314

& 0.2744 & 0.3031 &554.89& 74.53& 8.04& 548\\
 & w/ EC & 0.4603 & 0.2506 & 0.2814 &556.16& 62.47+\textcolor{red}{3.42}& 9.54+\textcolor{red}{0.08}& 548+\textcolor{red}{1}\\
 & w/o PD 
& 0.6134 & 0.616& 0.6147& 1055.36 & 97.15& 7.65& 1023\\
 & w/ PD  & 0.3787 & 0.6338 & 0.4741& 548.93 & 57.78+\textcolor{red}{0.99}& 8.80+\textcolor{red}{0.02}& 552+\textcolor{red}{2}\\
 
\bottomrule
\end{tabular}
\end{table}

\subsection{Breakdown Evaluation of \Sema}
\label{sec:exp:breakdown_analysis}

\begin{figure*}
    \begin{tabular}[t]{c}
        \centering
        \includegraphics[width=1.9\columnwidth]{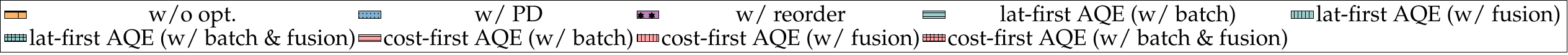}
	\end{tabular}
    \begin{tabular}[h]{c}
     \centering
        \subfigure[Query Latency (local)]{
				\includegraphics[width=0.68\columnwidth]{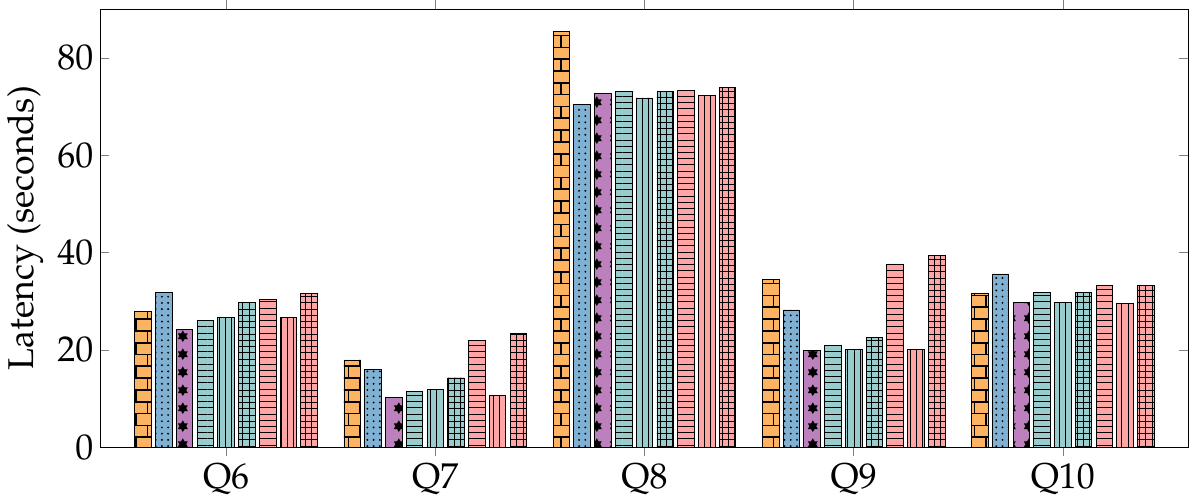}
			\label{fig:T_heatmap}
		} 
        \subfigure[Total Token Cost (local)] {
			\includegraphics[width=0.68\columnwidth]{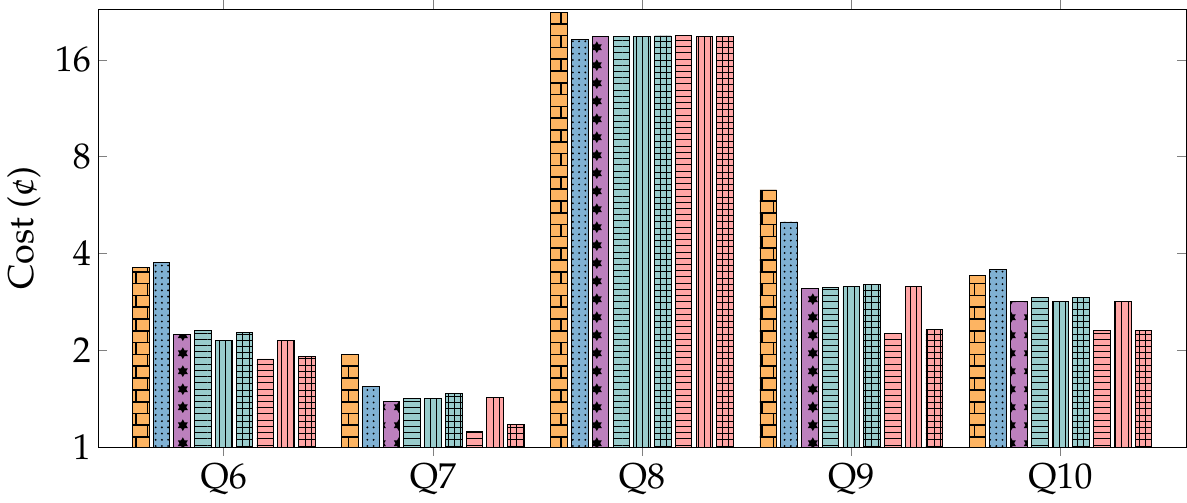}
			\label{fig:F_heatmap}
		} 
        \subfigure[F1 (local)] {
			\includegraphics[width=0.68\columnwidth]{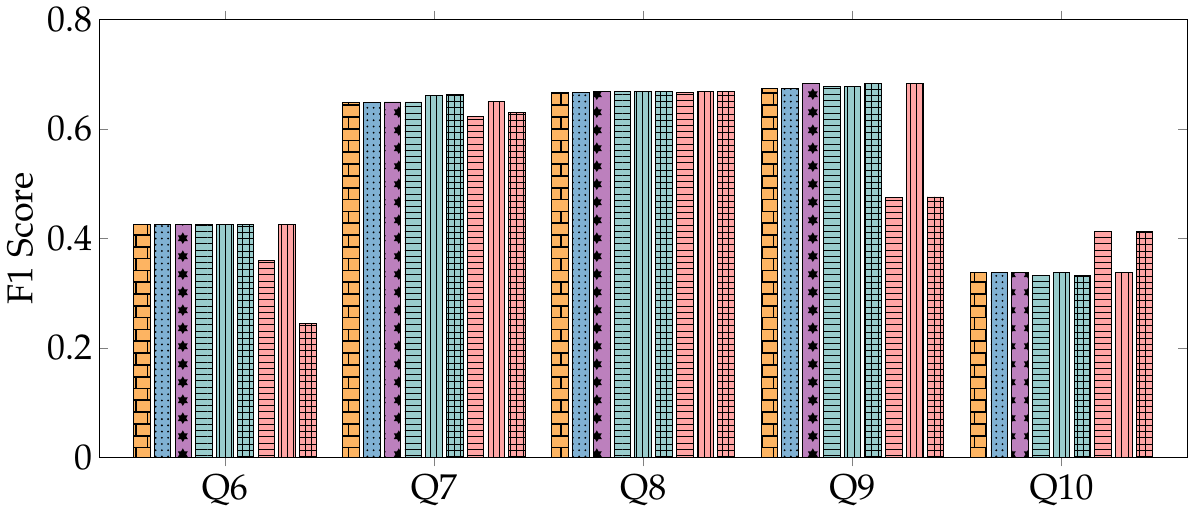}
			\label{fig:F_heatmap}
		}
	\end{tabular}
    \begin{tabular}[h]{c}
     \centering
        \subfigure[Query Latency (remote)]{
				\includegraphics[width=0.68\columnwidth]{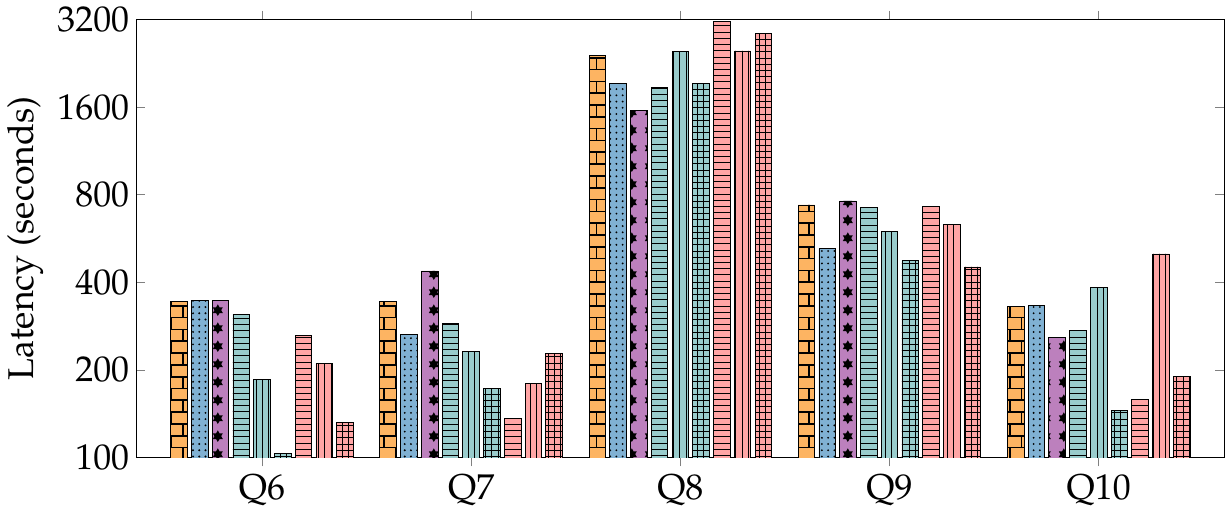}
			\label{fig:T_heatmap}
		} 
        \subfigure[Total Token Cost (remote)] {
			\includegraphics[width=0.68\columnwidth]{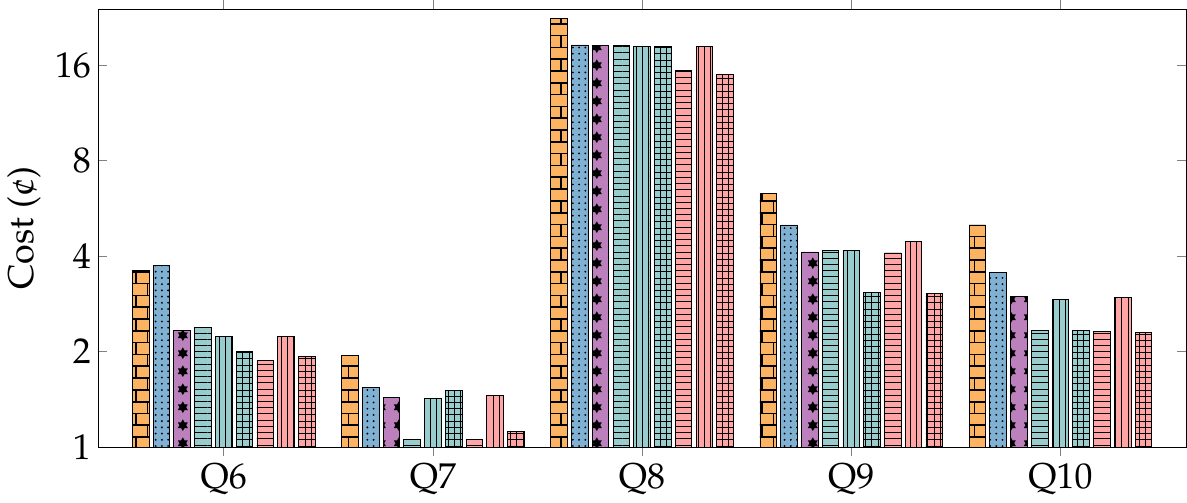}
			\label{fig:F_heatmap}
		} 
        \subfigure[F1 (remote)] {
			\includegraphics[width=0.68\columnwidth]{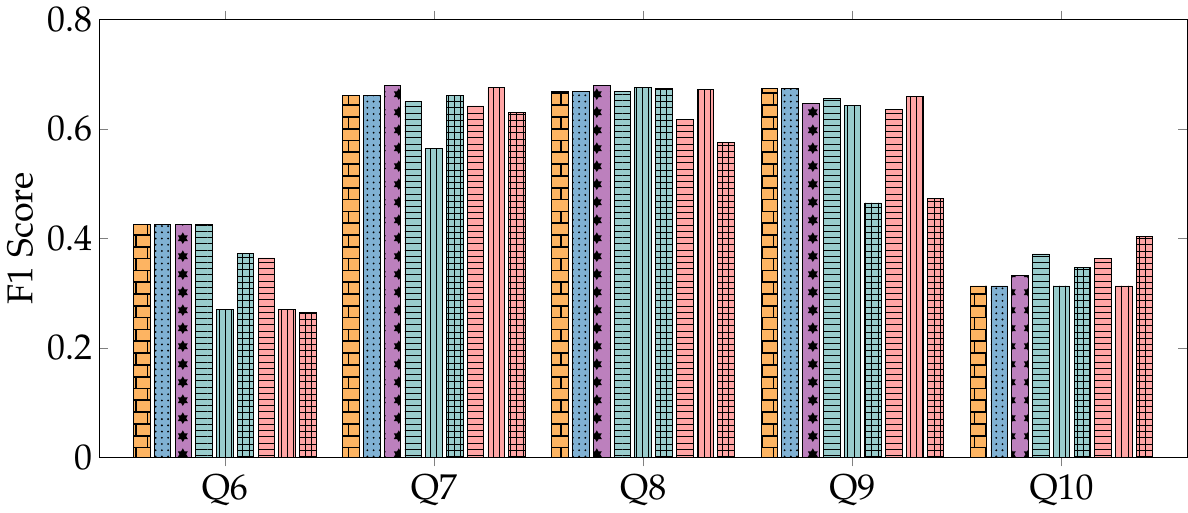}
			\label{fig:F_heatmap}
		}
	\end{tabular}
    \vspace{-4mm}
	\caption{Breakdown Evaluation of \Sema regarding Query Latency, Token Cost and F1 Under Local and Remote APIs}
        \label{fig:breakdown}
    
\end{figure*}

To investigate the effectiveness of our AQE framework of \Sema, we conduct experiments on queries Q6–Q10 in both remote and local scenarios, each involving multiple semantic filter predicates.
We set the fraction of data used for expression exploration and path exploration as $\delta_1 = \frac{1}{32}$ and $\delta_2 = \frac{3}{32}$, respectively, and the remaining $\frac{28}{32}$ data is processed by the selected path. 
We use local Gemma-3 12B for LLM inference.
We set the MCC threshold for \SemFilter fusion and $\tau_{\text{acc}}$ to 0.5 and 0.80, respectively.

Fig.~\ref{fig:breakdown} presents the comprehensive query-level metrics and latency breakdown for Q6–Q10 under various AQE configurations and two settings (local and remote). The three metrics exhibit substantial differences between the local and remote settings. Overall, the results highlight that batching introduces a cost–latency trade-off in local environments, while fusion can be latency-positive or latency-negative depending on the dataset; in remote deployments, both optimizations become effectively Pareto-improving, demonstrating environment-aware adaptive optimization.

In the local setting, both prompt batching and operator fusion consistently reduce total token cost. However, their latency impacts differ: prompt batching increases query latency, while the latency change introduced by operator fusion is dataset-dependent—it may either increase or decrease latency depending on the workload characteristics. The effectiveness of these techniques is also affected by the scheduler of vLLM, which is relatively more suitable for more and shorter LLM prompting than for fewer and longer LLM prompting. Therefore, the side-effect of prompt batching and operator fusion is that they may degrade the parallelism of \Sema, depending on the length of the prompts.
Consequently, under a cost-first AQE objective, the optimizer tends to favor fused and/or batched plans, achieving noticeable cost reductions at the expense of higher latency (primarily due to batching). Under a latency-first AQE objective, the optimizer avoids aggressive batching and applies fusion only when it is latency-beneficial for the given dataset, resulting in limited or inconsistent latency gains over the baseline. In the remote setting, the behavior changes substantially: because network round-trip overhead dominates execution time, reducing the number of remote calls via fusion and batching simultaneously lowers both latency and cost. As a result, both latency-first and cost-first AQE often select similar optimized plans and achieve consistent improvements in both metrics.

\stitle{Performance Characteristics: Local vs. Remote.}
The latency of LLM-based query execution is primarily governed by two factors: the number of LLM calls and the token length per call. However, the dominant factor shifts depending on the environment. For local LLM calls, execution is primarily constrained by local GPU computational power and memory size. In this scenario, throughput is relatively fixed. Consequently, latency is highly sensitive to token length per call, while the impact of call counts is less significant. While for remote LLM calls, performance is dominated by network overhead (e.g., round-trip time, handshakes, queue time) and API provider rate limits. In this context, the per-call overhead is substantial, making latency highly sensitive to the number of LLM calls, whereas moderate increases in token length per call have a negligible impact on total time.

\stitle{Trade-offs Under Multi-objective Strategies.}
When AQE is configured for multi-objective optimization, the results reveal clear and controllable trade-offs between latency and cost, consistent with our design intuition. Latency-first strategies consistently achieve the shortest query runtimes (e.g., 29.64s for Q6 and 51.94s for Q9), whereas cost-first strategies yield the lowest monetary costs (e.g., 2.21\textcent{}  for Q6 and 3.04\textcent{} for Q9).

\begin{figure}
\begin{lstlisting}{sql}
SELECT coins.id AS coin_id FROM coins 
WHERE category = 'coin' AND description is NOT NULL
AND s'The {description} shows that this cryptocurrency has a big supply amount.'
AND s'The {description} indicates that this cryptocurrency is still in circulation.'
AND s'It can be inferred from the {description} that the price of this cryptocurrency is rising.');
\end{lstlisting}
\vspace{-2ex}
\caption{Q6 for AQE case study}
\label{fig:aqe_qeury}
\end{figure}

\begin{table}[!t]
\begin{center}
\caption{Micro-execution results of candidate paths for Q6}
\vspace{-3mm}
\label{tab:aqe_path_statistics}
\footnotesize
\begin{tabular}{c lccc}
\toprule
Description & Path & Acc (\%) & Latency (s)  & Cost (\$) \\
\midrule
($[\sigma_3, \sigma_1, \sigma_2]$) & $p_1$ & 100.00 & 0.997 & 0.0008 \\
($[\sigma_3, \sigma_1, \sigma_2]$) w/ batch &  \(\widehat{p}_1\) & 94.04  & 1.936 & 0.0006 \\
$[\sigma_2 \oplus \sigma_3, \sigma_1]$ & \(p_2\) & 100.00 & 0.504 & 0.0007 \\
$[\sigma_2 \oplus \sigma_3, \sigma_1]$ w/ batch &\(\widehat{p}_2\) & 91.66  & 1.412 & 0.0005 \\
\bottomrule
\end{tabular}
\end{center}
\end{table}

\begin{table*}[h]
\centering
\footnotesize
\caption{Planning time (in millisecond), and breakdown time for phase 1-3 of AQE (t1, t2, t3 in seconds)}
\vspace{-3mm}
\label{table:planning_time}
\setlength{\tabcolsep}{3pt}
\begin{tabular}{l|cccc|cccc|cccc|cccc|cccc}
\toprule
\textbf{Metric} & \multicolumn{4}{c|}{\textbf{Q6}} & \multicolumn{4}{c|}{\textbf{Q7}} & \multicolumn{4}{c|}{\textbf{ Q8}} & \multicolumn{4}{c|}{\textbf{Q9}} & \multicolumn{4}{c}{\textbf{ Q10 }} \\ 
 - & $p_{\text{ref}}$ & $p_{\text{base}}$ & $p^\star_{\text{latency}}$  & $p^\star_{\text{cost}}$ & $p_{\text{ref}}$ & $p_{\text{base}}$ & $p^\star_{\text{latency}}$  & $p^\star_{\text{cost}}$ & $p_{\text{ref}}$ & $p_{\text{base}}$ & $p^\star_{\text{latency}}$  & $p^\star_{\text{cost}}$ & $p_{\text{ref}}$ & $p_{\text{base}}$ & $p^\star_{\text{latency}}$  & $p^\star_{\text{cost}}$ & $p_{\text{ref}}$ & $p_{\text{base}}$ & $p^\star_{\text{latency}}$  & $p^\star_{\text{cost}}$  \\ \midrule
\textbf{plan (ms)} & 30 & 30 & 29 & 30 & 54 & 51 & 52 & 55 & 51 & 55 & 49 & 49 & 54 & 54 & 56 & 50 & 49 & 46 & 56 & 54 \\
\textbf{t1 (s)}    & 0.00 & 0.57 & 0.57 & 0.57 & 0.00 & 0.32 & 0.33 & 0.34 & 0.00 & 0.59 & 0.62 & 0.65 & 0.00 & 0.61 & 0.60 & 0.64 & 0.00 & 1.30 & 1.35 & 1.32 \\
\textbf{t2 (s)}    & 0.00 & 0.00 & 3.45 & 4.86 & 0.00 & 0.00 & 1.79 & 1.81 & 0.00 & 0.00 & 3.93 & 5.72 & 0.00 & 0.00 & 1.42 & 1.45 & 0.00 & 0.00 & 3.24 & 3.24 \\
\textbf{t3 (s)}    & 0.00 & 24.60 & 25.61 & 30.85 & 0.00 & 41.49 & 40.86 & 71.81 & 0.00 & 70.87 & 70.24 & 120.05 & 0.00 & 39.86 & 39.30 & 54.82 & 0.00 & 28.75 & 27.29 & 26.60 \\ \bottomrule
\end{tabular}
\end{table*}

\stitle{Case Study of AQE.} We further illustrate AQE dynamics using Q6 (in Fig.~\ref{fig:aqe_qeury}) with local LLM calls, which retrieves cryptocurrencies with large supply ($\sigma_1$), active circulation ($\sigma_2$), and rising prices ($\sigma_3$).
In the Expression Exploration Phase, we independently evaluated three predicates on a $\frac{1}{32}$ subsample (27 rows), estimating reflectivities as \(s(\sigma_1) = 0.66\), \(s(\sigma_2) = 0.77\), and \(s(\sigma_3) = 0.22\). 
Correlation analysis on MCC reveals \(\mathrm{MCC}(\sigma_1,\sigma_2)=0.755\), \(\mathrm{MCC}(\sigma_1,\sigma_3)=0.377\).
There is a strong positive correlation between \(\sigma_1\) and \(\sigma_2\), indicating a potential of operator fusion, and \(\sigma_3\) is the most selective and relatively independent predicate.
 Guided by these statistics, \Sema generates four new execution paths, a base path $p_1$ with \SemFilter reordering ($[\sigma_3, \sigma_1, \sigma_2]$), a variant of base path $p_2$ with operator fusion ($[\sigma_2 \oplus \sigma_3, \sigma_1]$), and variants of $p_1$ and $p_2$ with prompt batching, denoted as $\widehat{p}_1$ and $\widehat{p}_2$, respectively. 
In the {Path Exploration Phase}, these candidate paths are micro‑executed on a $\frac{3}{32}$ subsample (84 rows here) to compare accuracy (against the base path), latency, and cost. The results of these paths are summarized in Table~\ref{tab:aqe_path_statistics}, where the `Acc' here represents the relative Acc. of the path to $p_1$.

Predicate fusion ($p_2$) substantially reduces token I/O and latency, achieving up to 49\% lower latency and 9\% lower cost compared to the base path ($p_1$) while retaining full accuracy. In contrast, batching does not always yield latency gains in this case. We observe that batched executions (\(\widehat{p}_1\) and \(\widehat{p}_1\)) sometimes incur longer delays than their non-batched counterparts. This difference is likely due to environmental constraints: while reducing LLM requests, batching increases per-request size and reduces concurrency, which lengthens end-to-end processing time in a local setting. Given an accuracy threshold of $\tau_{acc} = 0.80$, paths \(\widehat{p}_1\) and $p_2$ lie on the Pareto frontier. Under a latency-first strategy, \(p_2\) is selected as the optimal path \(p^\star\); under a cost-first strategy, \(\widehat{p}_1\) is preferred.

\stitle{Planning Time}.
We report the execution time of the three stages in AQE in Table \ref{table:planning_time}. Generally, the planning time is consistently small across all queries and plan variants. The end-to-end overhead is dominated by the final path exploitation phase ($t_3$), while the two exploration phases ($t_1$ and $t_2$) incur only modest cost. AQE’s planning/exploration cost remains lightweight relative to exploitation, confirming that the framework spends most of its time in exploitation rather than exploration.

\section{Related Work}
\label{sec:related}

\stitle{In-database Machine Learning.}
Equipping database systems with ML functionalities is a long-standing research topic.
Early studies~\cite{DBLP:journals/sigmod/LuoGGPJ18, DBLP:conf/dasfaa/ZhaoYRLH21} use relational algebra to represent linear algebra operations, where ML models can be expressed and executed in SQL within a DBMS. 
Another line is to implement UDFs and computation libraries over matrix and vector primitives. 
For example, MADlib~\cite{DBLP:journals/pvldb/HellersteinRSWFGNWFLK12} is an in-database library providing matrix and vector operations, and both UDFs and UDAs are employed to implement gradient descent~\cite{DBLP:journals/debu/BorkarBCRPCWR12, DBLP:conf/sigmod/FengKRR12}. 
To support a variety of statistical models, Monte Carlo database systems~\cite{DBLP:conf/sigmod/CaiVPAHJ13, DBLP:journals/tods/JampaniXWPJH11} have been designed, providing specialized UDFs for sampling in the database, thus enabling Bayesian learning through SQL queries. 
The concept of model-based views~\cite{DBLP:conf/sigmod/DeshpandeM06, DBLP:journals/pvldb/KocR11} has also been introduced to maintain ML models in DBMS following updates of the underlying training data.
Nowadays, many commercial and cloud database systems integrate ML capabilities, such as BigQuery ML~\cite{bigquery}, Amazon Redshift ML~\cite{redshift}, Oracle Data Mining~\cite{oracle}. 
More recently, DBMS injects deep learning models into UDFs to deal with complex analytical queries, where LLMs are either invoked in UDFs~\cite{liu2025optimizing, DBLP:journals/corr/abs-2504-01157}, or are used to generate the code of UDFs~\cite{DBLP:journals/pacmmod/ZhangSHKB25}.

\stitle{General ML-based Analytical System.}
The widespread adoption of ML has driven the development of general ML-based analytical systems.
Video database management systems (VDBMS)~\cite{DBLP:journals/pvldb/BangKCMA23, DBLP:journals/pvldb/RomeroHPKZK22, DBLP:journals/pvldb/KangBZ19, DBLP:journals/pvldb/KangGBHZ20, DBLP:journals/pvldb/KangMVBZ20, DBLP:journals/pvldb/XiaoZLWT023} are specifically designed for video data analysis, leveraging DL models as UDFs in DBMS for tasks such as image classification and object detection. 
To efficiently process video analytical queries, these systems incorporate a variety of optimization techniques, including end-to-end optimization of the preprocessing and query pipeline~\cite{DBLP:journals/pvldb/KangMVBZ20}, declarative query hints~\cite{DBLP:journals/pvldb/RomeroHPKZK22}, fine-grained model selection~\cite{DBLP:conf/sigmod/CaoSHAK22}, query-agnostic semantic index~\cite{DBLP:journals/pvldb/BangKCMA23}, and fast evaluation algorithms for specific operators~\cite{DBLP:journals/pvldb/KangBZ19, DBLP:journals/pvldb/BangKCMA23}.
Beyond video analysis, ML-powered probabilistic predicates~\cite{DBLP:journals/pvldb/DingAL22, DBLP:conf/sigmod/LuCKC18, DBLP:journals/pvldb/YangWHLLW22} have been applied in a broader range of applications. These studies construct lightweight proxy models and exploit the correlations between different predicates. 
For hybrid queries that combine ML and relational operators, RAVEN~\cite{DBLP:conf/sigmod/ParkSBSIK22} performs cross optimization of ML and relational operators, exploiting data statistics to select across multiple execution strategies. 
In addition, some systems also adopt adaptive query processing by dynamically  adjusting query plans and resource allocation, or choosing cheaper proxy models, following runtime statistics.
ABAE~\cite{DBLP:journals/pvldb/KangGBHSZ21}, InQuest~\cite{DBLP:journals/pvldb/RussoHK0Z23} specialize in optimizing aggregate queries with ML models by introducing sampling-based aggregate algorithms.

\stitle{LLM-based Semantic Query Processing.} LLMs have introduced a new paradigm in semantic query processing for both structured data~\cite{DBLP:journals/corr/abs-2504-01157, liu2025optimizing} and unstructured data~\cite{DBLP:conf/icde/WangF25, DBLP:journals/corr/abs-2410-12189, DBLP:journals/corr/abs-2409-00847}. Systems such as DocETL~\cite{DBLP:journals/corr/abs-2410-12189},  
ZenDB~\cite{DBLP:journals/corr/abs-2405-04674}, DocWanger~\cite{DBLP:journals/corr/abs-2504-14764}, 
Aryn~\cite{DBLP:journals/corr/abs-2409-00847} and Unify~\cite{DBLP:conf/icde/WangF25} leverage LLMs for analytics over unstructured text by semantic queries, supporting a broad spectrum of query interfaces including extended SQL dialects~\cite{DBLP:journals/corr/abs-2405-04674}, declarative operations~\cite{DBLP:journals/corr/abs-2410-12189} and natural language~\cite{DBLP:journals/corr/abs-2504-14764, DBLP:conf/icde/WangF25, DBLP:journals/corr/abs-2409-00847}. 
In these document analytical systems, LLMs also serve as optimization agents during  query evaluation, facilitating tasks such as query planning~\cite{DBLP:journals/corr/abs-2409-00847}, query rewriting, prompt refinement and operation decomposition~\cite{DBLP:journals/corr/abs-2504-14764}. 
UQE~\cite{dai2024uqe} extends SQL by allowing natural language join and selection predicates, which intrinsically supports semantic join and semantic filter by LLMs.  
As an LLM-powered analytical system, Palimpzest~\cite{liu2025palimpzest} provides logical and physical optimizations to reduce the query time and financial cost of LLM invocations.
Building on Palimpzest, Abacus~\cite{russo2025abacus} introduces a Pareto optimization algorithm and cost estimation for query planning. 
Lotus~\cite{patel2024lotus} formulates LLM-based semantic operations over tabular data and proposes optimized algorithms for costly semantic operations.
\cite{akillioglu2025} evaluates semantic query performance of PostgreSQL with pgAI and DuckDB with FlockMTL~\cite{DBLP:journals/corr/abs-2504-01157}, highlighting the ongoing challenges in enforcing structured outputs, batch processing, and query planning.
To the best of our knowledge, \Sema is the first system that supports native semantic operators and implements end-to-end optimization on top of DuckDB. 

\stitle{Adaptive Query Execution (AQE).}
Over the past two decades, AQE has progressed from fine-grained tuple routing to robust, domain-specific execution frameworks. The notion of continuous adaptivity was pioneered by Eddies~\cite{DBLP:conf/sigmod/HellersteinA00}, which replaced static operator trees with real-time tuple routing. This idea was later refined by CBR~  \cite{DBLP:conf/vldb/BizarroBDW05}, which demonstrated that routing based on data content outperforms simple average statistics. However, to mitigate the high overhead of tuple-level routing, subsequent research has shifted toward more practical, stage-based adaptation. For example, Polar~\cite{DBLP:journals/pvldb/JustenRFLTLBHZMB24} mitigates cardinality estimation errors by adaptively switching join orders through plans of least resistance, enabling robust execution without reconstructing the optimizer. More  recently, AQE has expanded to support the growing prevalence of ML operators in DBMSs. 
Aero~\cite{DBLP:journals/pacmmod/KakkarCSAK25} applies AQE to optimize ML-centric queries, dynamically adjusting the invocation of expensive UDFs and probabilistic models based on runtime data characteristics. 
In this work, due to the inherent unpredictability and uncertainty of the execution of LLM-based operators, \Sema employs AQE in the executor to explore a trade-off between cost, latency, and result quality.

\section{Conclusion}
\label{sec:conclusion}
In this work, we presented \Sema, a high-performance system that integrates semantic operators for LLM-based semantic query processing in database systems. 
To optimize such semantic queries, \Sema introduces natural language  expression  optimizations for semantic operators at the optimizer stage and incorporates runtime optimization strategies at the executor stage. Considering the inherent uncertainty of LLMs, \Sema adopts an adaptive query execution framework to achieve a balanced trade-off among quality, cost, and latency during execution. Our results demonstrate the potential of unifying semantic reasoning with traditional data management for more adaptive and intelligent query processing. 
An executable version of the system \Sema is released in \url{https://github.com/BITQiKangK/SemaSystem.git}.

\bibliographystyle{ACM-Reference-Format}
\bibliography{ref}

\clearpage
\appendix
\section{Exemplar Queries}
\label{sec:appendix:example}

\begin{figure}[H]
\centering
\begin{minipage}{0.95\columnwidth}
\begin{lstlisting}{sql}
-- Extract keywords from a content field 
SELECT id, s'Extract the main keywords from {content}' AS extracted_keywords 
FROM documents; 
-- Create a semantic summary from multiple fields 
SELECT s'A summary of the event based on {title} and {description}' 
AS event_summary FROM news_articles;
\end{lstlisting}
\caption{Example of Semantic Projection}
\label{fig:semantic_projection}
\end{minipage}
\end{figure}

\begin{figure}[H]
\centering
\begin{minipage}{0.95\columnwidth}
\begin{lstlisting}{sql}
-- Filter rows based on the sentiment of a specific field 
SELECT * FROM product_reviews 
WHERE s'The {review_text} expresses positive sentiment';
\end{lstlisting}
\caption{Example of Semantic Filter}
\label{fig:semantic_filter}
\end{minipage}
\end{figure}

\begin{figure}[H]
\centering
\begin{minipage}{0.95\columnwidth}
\begin{lstlisting}{sql}
-- Analyze the language style of the conference paper title
WITH title_style AS (
    SELECT Title, s'Analyze the language style features of paper title {Title}. 
    Describe the key characteristics such as: formality level, 
    technical terminology usage, length, structure, and word choice.' 
    AS style_features
    FROM Paper WHERE ConferenceId != 0 AND Title IS NOT NULL
)
SELECT sem_agg(s'Based on these conference paper titles {Title} 
and their style analyses {style_features}, summarize the overall 
language style and terminology characteristics of conference paper titles.')
FROM title_style;
\end{lstlisting}
\caption{Example of Semantic Aggregate}
\label{fig:semantic_aggregate}
\end{minipage}
\end{figure}

\begin{figure}[H]
\centering
\begin{minipage}{0.95\columnwidth}
\begin{lstlisting}{sql}
-- Join two tables based on semantic similarity of their text fields 
SELECT * FROM news_articles a JOIN blog_posts b 
ON s'the topic of {a.headline} is related to {b.title}';
-- A complex hybrid join with both syntactic and semantic conditions 
SELECT e.employee_id, eh.manager_id 
FROM employees e INNER JOIN employee_hierarchy eh 
ON e.manager_id = eh.employee_id 
AND s'the {e.title} is a technical role and the hierarchy level {eh.level} < 3';
\end{lstlisting}
\caption{Example of Semantic Join}
\label{fig:semantic_join}
\end{minipage}
\end{figure}

\section{Algorithms}
\label{sec:appendix:algorithm}

\begin{algorithm}[t]
\small
\caption{Natural Language Expression Reduction}
\label{alg:nle-preprocessing}
\SetKwInOut{Input}{Input} \SetKwInOut{Output}{Output}
\Input{userInputNLE}
\Output{processedNLE}
reductionStatus $\leftarrow$ \textsc{AssessDbReduction}(userInputNLE) \\
\If{reductionStatus = "not\_reducible"}{
  \Return processedNLE $\leftarrow$ cleanedNLE
}
processedNLE $\leftarrow$ \textsc{ExpressionReduction}(userInputNLE) \\
\Return processedNLE
\end{algorithm}

\begin{algorithm}[t]
\small
\caption{Expression Exploration and Pairwise Similarity Estimation}
\label{alg:expr-explore}
\SetKwInOut{Input}{Input} \SetKwInOut{Output}{Output}
\Input{$\Sigma = \{\sigma_1,\dots,\sigma_n\}$, sample chunk $\mathcal{D}_1$}
\Output{Selectivities $s$, result vectors $R$, similarity matrix $\mathrm{MCC}$}
\For{$i \gets 1$ to $n$}{
  $R[i] \leftarrow \textsc{ExecutePredicate}(\sigma_i, \mathcal{D}_1)$ \;
  $S[i] \leftarrow \frac{\textsc{CountTrue}(R[i])}{|\mathcal{D}_1|}$ \;
}
\For{$i \gets 1$ to $n$}{
  \For{$j \gets i+1$ to $n$}{
    $(TP,TN,FP,FN) \leftarrow \textsc{ConfusionCounts}(R[i],R[j])$ \;
    $\mathrm{MCC}[i][j] \leftarrow \textsc{Matthews}(TP,TN,FP,FN)$ \;
  }
}
\Return $S, R, \mathrm{MCC}$
\end{algorithm}

\begin{algorithm}[t]
\small
\caption{Candidate Path Generation with Dynamic Pruning}
\label{alg:path-gen}
\SetKwInOut{Input}{Input} \SetKwInOut{Output}{Output}
\Input{$S$, $\mathrm{MCC}$, $\Sigma$, $\mu_{\text{thresh}}$}
\Output{Candidate paths $\mathcal{P}$}
$p_{\text{sorted}} \leftarrow \textsc{SortAscendingBySelectivity}(\Sigma,S)$ \;
$\mathcal{P} \leftarrow \{\textsc{SeqPath}(p_{\text{sorted}})\}$ \;
$\mathcal{F} \leftarrow \emptyset$ \;
\For{$i \gets 1$ to $n$}{
  \For{$j \gets i+1$ to $n$}{
    \If{$\mathrm{MCC}[i][j] > \mu_{\text{thresh}}$}{
      $\mathcal{F} \leftarrow \mathcal{F} \cup \{(\sigma_i,\sigma_j)\}$ \;
    }
  }
}
\For{each $(\sigma_a,\sigma_b) \in \mathcal{F}$}{
  $\sigma_a \oplus \sigma_b \leftarrow \textsc{Fuse}(\sigma_a,\sigma_b)$ \;
  $\mathcal{P} \leftarrow \mathcal{P} \cup \{\textsc{SeqPath}(\textsc{ReplaceOnce}(p_{\text{sorted}},\{\sigma_a,\sigma_b\}\to\sigma_a\oplus\sigma_b))\}$ \;
}
\Return $\mathcal{P}$
\end{algorithm}

Algorithm \ref{alg:expr-explore} evaluates each predicate ($\sigma_i \in \Sigma$ ) on the sampled data chunk ($\mathcal{D}_1$) to produce a Boolean result vector $R[i]$, from which the selectivity $S[i]$ is estimated as the fraction of tuples satisfying the predicate (line 1-3). Using these result vectors, the algorithm then computes pairwise predicate similarity by deriving a confusion matrix for each predicate pair and applying the Matthews Correlation Coefficient (MCC) to populate the similarity matrix (line 4-7).

Algorithm \ref{alg:path-gen} generates a set of candidate execution paths with dynamic pruning guided by predicate correlations. Given a set of predicates $\Sigma$, the algorithm first sorts all predicates in ascending order of selectivity to construct a baseline sequential execution path (line 1), which is included in the candidate set $\mathcal{P}$ (line 2). It then examines pairwise predicate correlations using a precomputed Matthews Correlation Coefficient (MCC) matrix. It retains only those predicate pairs whose correlation exceeds a threshold $\mu_{\text{thresh}}$ in $\mathcal{F}$ (line 4-7), effectively pruning weakly related combinations. For each retained pair $(\sigma_i, \sigma_j)$, the algorithm applies a fusion function to form a fused predicate $\sigma_i \oplus \sigma_j$ and generates a new candidate path by replacing exactly one occurrence of the original predicates in the sorted sequence with the fused predicate, while preserving the relative order of all remaining predicates (line 8-10). All generated paths are collected into $\mathcal{P}$ and treated as candidate paths for the subsequent path exploration phase.

\begin{algorithm}[t]
\small
\caption{Path Exploration and Best-Path Selection}
\label{alg:path-select}
\SetKwInOut{Input}{Input} \SetKwInOut{Output}{Output}
\Input{Paths $\mathcal{P}$, sample chunk $\mathcal{D}_2$, threshold $\tau_{\text{acc}}$}
\Output{Best path $p^\star$}
Execute reference path to get ground truth $G$ \;
\For{each $p \in \mathcal{P}$}{
  $(\text{lat},\text{cost},\text{sel}) \leftarrow \textsc{Run}(p,\mathcal{D}_2)$ \;
  $\text{acc} \leftarrow \textsc{Accuracy}(\text{sel}, G)$ \;
}
$\mathcal{F} \leftarrow \textsc{ParetoFrontier}(\mathcal{P}, \text{lat}, \text{cost})$ \;
$p^\star \leftarrow \arg\min_{p \in \mathcal{F}} \mathcal{M}[p].\text{lat}$ \;
\Return $p^\star$
\end{algorithm}

Algorithm \ref{alg:path-select} explores the candidate execution paths and selects the best path for the exploitation phase. Given a set of candidate paths $\mathcal{P}$ and a sampled data chunk $\mathcal{D}_2$, the algorithm first executes the reference path $p_{\text{ref}}$ to obtain ground-truth results (line 1). Each candidate path is then executed on $\mathcal{D}_2$ to measure its latency, execution cost, and selection output, from which accuracy is computed against the ground truth (line 2-4). Based on the collected latency and cost metrics, the algorithm constructs a Pareto frontier to retain only non-dominated paths (line 5). Among these Pareto-optimal candidates, the path with the minimum latency or minimum cost is selected as the final best path $p^\star$ and returned for execution (line 6-7).

\section{Dataset \& Benchmark Queries}
\label{sec:appendix:benchmark}

\begin{table*}[htbp]
\footnotesize
\setlength{\tabcolsep}{4.5pt} 
\renewcommand{\arraystretch}{1.15}
\caption{Datasets and Queries Used in the Experiments}
\vspace{-2ex}
\label{tab:datasets}
\centering
\begin{tabularx}{\textwidth}{
    @{}l
    c
    >{\raggedright\arraybackslash}p{6cm}
    >{\raggedright\arraybackslash}p{4.2cm}
    >{\centering\arraybackslash}p{1.8cm}
    @{}
}
\toprule
\textbf{Dataset} & \textbf{\# Tables} & \textbf{Table (\# Rows)} & \textbf{Attributes} & \textbf{Queries} \\
\midrule
Appstore & 2 & \kw{user\_review} (10,840), \kw{playstore} (64,286) & \kw{user\_review.Translated\_Review} & Q7, Q8 \\

Chicago\_Crime & 1 & \kw{IUCR} (401) & \kw{primary\_description}, \kw{secondary\_description}, \kw{index\_code} & Q3 \\

Superstore & 1 & \kw{product} (5,298) & \kw{productName}, \kw{category} & Q2 \\

Food\_Inspection & 1 & \kw{violations} (36,050) & \kw{description}, \kw{risk\_category} & Q4 \\

Food\_Inspection\_2 & 1 & \kw{violation} (525,709) & \kw{inspector\_comment, fine} & Q5 \\

Social\_Media & 3 & \kw{twitter} (99,901), \kw{user} (99,260), \kw{location} (6,211) & \kw{twitter.text} & Q1, Q9, Q16–Q18 \\

Movies\_4 & 7 & \kw{movie} (4,627), \kw{country} (88), \kw{production\_country} (6,436), \kw{genre} (20), \kw{movie\_genres} (12,160), \kw{language} (88), \kw{movie\_languages} (11,740) & \kw{movie.overview} & Q10, Q13–Q15 \\

CoinMarketCap & 2 & \kw{coins} (8,927), \kw{historical} (4,441,972) & \kw{coins.description} & Q6, Q11, Q12 \\

Authors & 5 & \texttt{Paper} (2,254,920), \kw{Journal} (15,151), \kw{Conference} (4,545), \kw{Author} (247,030), \kw{PaperAuthor} (2,315,574) & \kw{Paper.Keyword, Paper.Title} & Q19, Q20 \\
\bottomrule
\end{tabularx}
\end{table*}

\begin{table}[H]
\centering
\caption{Profile of Query Set}
\label{tab:query_classification}
\begin{tabular}{@{}llc@{}}
\toprule
\textbf{Category} & \textbf{Subcategory} & \textbf{Query ID} \\ \midrule
Single \SemFilter/\SemProj & Single Operators & Q1--Q5 \\
Consecutive \SemFilter & Filter $\to$ Filter & Q6--Q10 \\
\SemProj-First & Proj. $\to$ Filter & Q11--Q13 \\
 & Proj. $\to$ Proj. & Q14--Q15 \\
\SemFilter-First & Filter $\to$ Proj. & Q16--Q18 \\
 & Filter $\to$ Aggregate & Q19--Q20 \\ \bottomrule
\end{tabular}
\end{table}

\begin{figure}[H]
\centering
\begin{minipage}{0.45\textwidth}
\centering
\begin{lstlisting}{sql}
SELECT id FROM (
    SELECT TweetID AS id, sentiment, text FROM twitter LIMIT 32768
) WHERE s'The sentiment label rules are:
- sentiment > 0 -> Positive
- sentiment = 0 -> Neutral
- sentiment < 0 -> Negative
Task: Judge whether the {sentiment} label is consistent with the {text} content.');
\end{lstlisting}
\end{minipage}
\vspace{-5pt}
\caption{Q1: Sentiment analysis}
\label{fig:q1}
\vspace{4pt}
\end{figure}

\begin{figure}[H]
\centering
\begin{minipage}{0.45\textwidth}
\centering
\begin{lstlisting}{sql}
SELECT s'Classify the given {"Product Name"} into exactly one of these categories:
[Furniture, Office Supplies, Technology].
Output only the category name, with no explanations, no quotes, and no extra words.
Example:
Input: Hon Metal Bookcases, Black Output: Furniture
Input: AT&T CL2909 Output: Technology
Input: Belkin F9S820V06 8 Outlet Surge Output: Office Supplies' AS predicted_category FROM product;
\end{lstlisting}
\end{minipage}
\vspace{-5pt}
\caption{Q2: Classify product into one of three categories}
\vspace{4pt}
\end{figure}

\begin{figure}[H]
\centering
\begin{minipage}{0.45\textwidth}
\centering
\begin{lstlisting}{sql}
SELECT iucr_no FROM IUCR 
WHERE s'The index_code label rules are:
- index_code = I -> Indexed (severe, such as murder, rape, arson, and robbery)
- index_code = N -> Non-Indexed (less serve, such as vandalism, weapons violations, and peace disturbance)
Task: Determine whether the crime''s {primary_description} and {secondary_description} indicates this crime is consisted with {index_code}');
\end{lstlisting}
\end{minipage}
\vspace{-5pt}
\caption{Q3: Determine whether the crime's description is consistent with the index code}
\vspace{4pt}
\end{figure}

\begin{figure}[H]
\centering
\begin{minipage}{0.45\textwidth}
\centering
\begin{lstlisting}{sql}
SELECT s'Task: Classify the given violation {description} into exactly one of these categories:
- Low Risk
- Moderate Risk
- High Risk
Output only the risk category, with no explanations, no quotes, and no extra words.
Example:
Input: Inadequate and inaccessible handwashing facilities. Output: Moderate Risk
Input: Unclean or degraded floors, walls, or ceilings.
Output: Low Risk
Input: Unclean or unsanitary food contact surfaces.
Output: High Risk' AS predicated_category, risk_category AS ground_truth FROM violations;
\end{lstlisting}
\end{minipage}
\vspace{-5pt}
\caption{Q4: Determine the risk degree according to the violation description}
\vspace{4pt}
\end{figure}

\begin{figure}[H]
\centering
\begin{minipage}{0.45\textwidth}
\centering
\begin{lstlisting}{sql}
SELECT inspection_id FROM (
    SELECT distinct inspection_id,inspector_comment, fine FROM violation
    WHERE inspector_comment IS NOT NULL
      AND fine IN [100, 250, 500]
    LIMIT 16384
)
WHERE s'The fine for more serious food safety problems will be higher.
The fine label rules are:
- fine = 100 -> Minor
- fine = 250 -> Serious
- fine = 500 -> Critical
Fine example:
1. {inspector_comment}: REPAIR THE DRAIN STOPPER FOR THE RIGHT COMPARTMENT OF THE 3 COMPARTMENT SINK SO THAT IT CAN BE USED TO HOLD WATER. {fine}: 100
2. {inspector_comment}: NO CHEMICAL TEST STRIPS ON PREMISES-MUST PROVIDE. SERIOUS VIOLATION 7-38-030. {fine}: 250
3. {inspector_comment}: FOUND WALK IN COOLER TEMPERATURE AT 45.3F. CRITICAL VIOLATION.7-38-005( A)  {fine}: 500
Task: Determine whether the {inspector_comment} is consistent with the {fine}. Output true or false only.';
\end{lstlisting}
\end{minipage}
\vspace{-5pt}
\caption{Q5: Determine the risk degree according to the violation description}
\vspace{4pt}
\end{figure}

\begin{figure}[H]
\centering
\begin{minipage}{0.45\textwidth}
\centering
\begin{lstlisting}{sql}
SELECT coins.id AS coin_id FROM coins 
WHERE category = 'coin' AND description is NOT NULL
AND s'The {description} shows that this cryptocurrency has a big supply amount.'
AND s'The {description} indicates that this cryptocurrency is still in circulation.'
AND s'It can be inferred from the {description} that the price of this cryptocurrency is rising.'
\end{lstlisting}
\end{minipage}
\vspace{-5pt}
\caption{Q6: Find cryptocurrencies that are still in circulation and have a huge supply amount with a rising price}
\vspace{4pt}
\end{figure}

\begin{figure}[H]
\centering
\begin{minipage}{0.45\textwidth}
\centering
\begin{lstlisting}{sql}
SELECT ur.id AS id 
FROM user_reviews ur
JOIN playstore p ON ur.App = p.App 
WHERE p.Category = 'TOOLS'
AND s'{ur.Translated_Review} is a complete sentence and has informative content.'
AND s'{ur.Translated_Review} is a positive user review.'
AND s'{ur.Translated_Review} shows that this app is a useful tool.'
\end{lstlisting}
\end{minipage}
\vspace{-5pt}
\caption{Q7: Find positive, complete, and informative user reviews of TOOLS apps that indicate the app is a useful tool}
\vspace{4pt}
\end{figure}

\begin{figure}[H]
\centering
\begin{minipage}{0.45\textwidth}
\centering
\begin{lstlisting}{sql}
SELECT ur.id AS id
FROM user_reviews ur
JOIN playstore p ON ur.App = p.App  
WHERE p.Category = 'GAME'
AND s'{Translated_review} is a complete sentence and has informative content.'
AND s'{Translated_review} is a positive user review.'
AND s'{Translated_Review} shows that users have a good experience and fun with this app';
\end{lstlisting}
\end{minipage}
\vspace{-5pt}
\caption{Q8: Find positive, complete, and informative user reviews of GAME apps that indicate users have a good experience and fun with the app}
\vspace{4pt}
\end{figure}

\begin{figure}[H]
\centering
\begin{minipage}{0.45\textwidth}
\centering
\begin{lstlisting}{sql}
SELECT ur.id AS id
FROM user_reviews ur
JOIN playstore p ON ur.App = p.App
WHERE p.Category = 'FAMILY'
AND s'{Translated_Review} is a complete sentence and has informative content.'
AND s'{Translated_Review} is a positive user review.'
AND s'{Translated_Review} shows that this app is suitable for kids.';
\end{lstlisting}
\end{minipage}
\vspace{-5pt}
\caption{Q9: Find positive, complete, and informative user reviews of FAMILY apps that indicate the app is suitable for kids}
\vspace{4pt}
\end{figure}

\begin{figure}[H]
\centering
\begin{minipage}{0.45\textwidth}
\centering
\begin{lstlisting}{sql}
SELECT m.movie_id 
FROM movie m
WHERE s'The {m.overview} suggests that the movie is an action movie.'
AND s'The {m.overview} shows that the movie focuses on a central role.'
AND s'The {m.overview} shows that the movie is based on historical events';
\end{lstlisting}
\end{minipage}
\vspace{-5pt}
\caption{Q10: Find action movies that focus on a central role and are based on historical events}
\vspace{4pt}
\end{figure}

\begin{figure}[H]
\centering
\begin{minipage}{0.45\textwidth}
\centering
\begin{lstlisting}{sql}
SELECT coin_id FROM 
(SELECT s'According to {c.description}, introduce the situation of cryptocurrency circulation' AS circulation, c.id AS coin_id
    FROM coins c JOIN historical h ON h.coin_id = c.id
    AND h.date = '2019-01-09'
    AND c.status IN ('untracked', 'inactive')
    AND c.description IS NOT NULL)
WHERE s'{circulation} indicates cryptocurrency is under development or in testnet.';
\end{lstlisting}
\end{minipage}
\vspace{-5pt}
\caption{Q11: Find cryptocurrencies that were inactive or untracked on Jan 9, 2019, had missing price data, and whose description suggests they were still under development or operating on a testnet}
\vspace{4pt}
\end{figure}

\begin{figure}[H]
\centering
\begin{minipage}{0.45\textwidth}
\centering
\begin{lstlisting}{sql}
SELECT coin_id FROM 
(SELECT s'From the description {description}, extract information about the underlying technology of cryptocurrency, focusing on whether it mentions a proprietary blockchain, a native consensus mechanism, or an independent network. Provide a concise summary of these technological characteristics.' AS tech_features,
    coins.id AS coin_id
    FROM coins c JOIN historical h ON h.coin_id = c.id
    WHERE c.platform_id IS NOT NULL AND c.description IS NOT NULL
    AND h.circulating_supply IS NOT NULL
    AND h.total_supply IS NOT NULL
    AND h.date = '2019-01-10'
    )
WHERE s'Based on the technological characteristics {tech_features}, determine whether cryptocurrency has its own independent blockchain or consensus mechanism';
\end{lstlisting}
\end{minipage}
\vspace{-5pt}
\caption{Q12: Find coins whose summarized technological features indicate the cryptocurrency has its own blockchain or consensus mechanism}
\vspace{4pt}
\end{figure}

\begin{figure}[H]
\centering
\begin{minipage}{0.45\textwidth}
\centering
\begin{lstlisting}{sql}
SELECT movie_id FROM 
(SELECT s'Summarize the geographic and temporal background of the movie according to {overview}.' AS background, m.movie_id AS movie_id
    FROM movie m
    SEMI JOIN (
      SELECT mg.movie_id
      FROM movie_genres mg JOIN genre g ON mg.genre_id = g.genre_id
      WHERE g.genre_name IN ('Thriller', 'Comedy', 'Western')
    ) mg ON m.movie_id = mg.movie_id
    WHERE revenue > budget
      AND length(overview) < 255
 )
WHERE s'{background} indicates the story is set in modern times rather than historical periods.';
\end{lstlisting}
\end{minipage}
\vspace{-5pt}
\caption{Q13: Find profitable movies whose background indicates the story is set in modern times rather than historical periods with limited genres}
\vspace{4pt}
\end{figure}

\begin{figure}[H]
\centering
\begin{minipage}{0.45\textwidth}
\centering
\begin{lstlisting}{sql}
SELECT movie_id, s'Classify {plot} into one of the following categories:
[Sci-Fi, Crime, Thriller, Romance, Drama, Horror, Comedy, War, Fantasy, Family, Animation, Biography, Other]
Output only the category that best fits the movie, without any other text or reasoning content.' AS category FROM
(SELECT s'Summarize the main plot of {overview}' AS plot, movie_id
    FROM movie m
    SEMI JOIN (
        SELECT ml.movie_id
        FROM movie_languages ml
        JOIN language l ON ml.language_id = l.language_id AND l.language_name = 'English'
    ) ml ON m.movie_id = ml.movie_id AND movie_status = 'Released' AND popularity > 10 AND length(overview) < 255 AND length(overview) > 0
);
\end{lstlisting}
\end{minipage}
\vspace{-5pt}
\caption{Q14: Summarize the main plot of English-released popular movies and classify plots into genres}
\vspace{4pt}
\end{figure}

\begin{figure}[H]
\centering
\begin{minipage}{0.45\textwidth}
\centering
\begin{lstlisting}{sql}
SELECT movie_id,
s'This is a multi-label classification task. Given the {conflict}, identify what the main character is in conflict with, and classify it in the format "character vs X", where X is one or more of: [Character, Self, Society, Nature, Technology, Machine, Supernatural, Fate, Other].
Output only X, as a comma-separated list (e.g., Self, Society).
Do not include explanations or reasoning.' AS categories
FROM (SELECT movie_id, s'Extract all conflicts from movie {overview} as one sentence.' AS conflict
    FROM movie
    SEMI JOIN
        (SELECT pc.movie_id 
         FROM production_country pc JOIN country c
         ON pc.country_id = c.country_id AND c.country_iso_code = 'US')
    pc ON movie.movie_id = pc.movie_id AND (movie.release_date BETWEEN '2000-01-01' AND '2015-12-31') AND movie.vote_average > 6.0 AND length(overview) < 255;
\end{lstlisting}
\end{minipage}
\vspace{-5pt}
\caption{Q15: Extract all conflicts from high-rated movies between 2000 and 2015 and classify conflicts}
\vspace{4pt}
\end{figure}

\begin{figure}[H]
\centering
\begin{minipage}{0.45\textwidth}
\centering
\begin{lstlisting}{sql}
SELECT s'From the job posting tweet {t.text}, extract the specific job titles and required skills mentioned. Return them as a comma-separated list (e.g., Software Engineer, Python, AWS, Machine Learning). Focus on concrete job titles and technical skills' AS info,t.TweetID AS twitter_id
FROM twitter t JOIN user u
On t.UserID = u.UserID
WHERE user.Gender = 'Female'
AND t.Sentiment = 0.0
AND t.TweetID IS NOT NULL
AND Klout > 50
AND s'Tweets with content as {text} are about job postings, recruitment announcements, hiring notices, or career opportunities related to AWS or cloud computing (has contents like hiring, job opening, career opportunity, we are looking for, join our team, etc.)';
\end{lstlisting}
\end{minipage}
\vspace{-5pt}
\caption{Q16: Filter neutral tweets which are posted by females and related to recruitment and extract concrete job titles and technical skills.}
\vspace{4pt}
\end{figure}

\begin{figure}[H]
\centering
\begin{minipage}{0.45\textwidth}
\centering
\begin{lstlisting}{sql}
SELECT s'Based on the tweet content :"{text}", identify the main type of security or privacy concern being expressed. Classify text like data breach, unauthorized access, encryption weakness, compliance violation, identity theft,
service vulnerability, configuration error, monitoring concerns, or other specific security/privacy concern types.' AS type,
twitter.TweetID AS twitter_id
FROM twitter 
WHERE Sentiment < -2.5 AND Klout > 20
AND t.TweetID IS NOT NULL
AND s'Tweet "{text}" discusses security concerns, privacy issues, data protection, cybersecurity threats, authentication problems, or confidentiality matters related to AWS services';
\end{lstlisting}
\end{minipage}
\vspace{-5pt}
\caption{Q17:Find negative tweets with high influence that are related to potential risks of AWS services and identify the main type of security or privacy concern}
\vspace{4pt}
\end{figure}

\begin{figure}[H]
\centering
\begin{minipage}{0.45\textwidth}
\centering
\begin{lstlisting}{sql}
SELECT s'From this tweet "{t.text}", extract the main topic or subject being discussed about programming/development. Return a concise topic label like performance optimization, learning resources, job opportunities, technical issues, new features, etc.'
twitter.TweetID AS twitter_id
FROM twitter t JOIN location l
On t.LocationID = l.LocationID
WHERE l.Country IN ('United States', 'Canada', 'United Kingdom')
AND Sentiment >= 3.0 AND Reach>1000
AND t.TweetID IS NOT NULL
s'Filter tweets "{t.text}" that mention Amazon Web Services or cloud services, including specific AWS products or general cloud computing issues';
\end{lstlisting}
\end{minipage}
\vspace{-5pt}
\caption{Q18: Filter tweets related to AWS products and extract discussed topics or subjects}
\vspace{4pt}
\end{figure}

\begin{figure}[H]
\centering
\begin{minipage}{0.45\textwidth}
\centering
\begin{lstlisting}{sql}
SELECT sem_agg('Summarize the main research theme reflected by the following keywords: {p.Keyword}. Respond with a concise, high-level scientific topic',Keyword) FROM Paper p
JOIN Journal j
ON j.Id= P.JournalId
WHERE p.Year BETWEEN 2000 AND 2011
AND j.ShortName IN ('TODS', 'VLDB', 'TKDE', 'JDM', 'DKE')
AND s'Based on the keywords "{p.Keyword}", determine if this paper is related to database systems, data management, or query processing';
\end{lstlisting}
\end{minipage}
\vspace{-5pt}
\caption{Q19: Find papers of Database journals and related to database systems, data management, or query processing, and aggregate these papers' keywords to find main research themes}
\vspace{4pt}
\end{figure}

\begin{figure}[H]
\centering
\begin{minipage}{0.45\textwidth}
\centering
\begin{lstlisting}{sql}
SELECT ConferenceId, sem_agg('Identify and summarize the major interdisciplinary research trends from the following keywords: {p.Keyword}. Return only one sentence.',Keyword) AS trends
FROM paper p 
JOIN Conference c ON p.ConferenceId = c.Id
SEMI JOIN (
SELECT DISTINCT pa.PaperId
  FROM PaperAuthor pa
  WHERE pa.AuthorId IS NOT NULL
    AND (LOWER(COALESCE(pa.Affiliation,'')) LIKE '%institute%'
    OR LOWER(pa.Affiliation) LIKE '%laboratory%'
    OR LOWER(pa.Affiliation) LIKE '%research center%')
) filtered_papers
ON p.Id = filtered_papers.PaperId
AND p.Keyword IS NOT NULL
AND length(p.Keyword) > 0
WHERE s'Determine whether the paper with keyword as {Keyword} is related to interdisciplinary methods.'
GROUP BY ConferenceId
HAVING COUNT(*) >= 3 AND trends IS NOT NULL
\end{lstlisting}
\end{minipage}
\vspace{-5pt}
\caption{Q20: Find papers whose authors are in institute, laboratory, or research center and identify interdisciplinary research trends}
\label{fig:q20}
\vspace{4pt}
\end{figure}

\begin{figure}[H]
\centering
\begin{minipage}{0.45\textwidth}
\centering
\begin{lstlisting}{sql}
-- Note: Introduces redundant expressions - repeated field references ({primary_description}, {secondary_description}, {index_code} each appear twice), redundant modifiers ("after reviewing...again for consistency"), and complex sentence structures.
SELECT COUNT(*) FROM chicago_crime.IUCR 
WHERE s'The index_code label rules are:\n\
- index_code = I -> Indexed (severe, such as murder, rape, arson, and robbery)\n\
- index_code = N -> Non-Indexed (less severe, such as vandalism, weapons violations, and peace disturbance)\n\
Based on these rules, the crime with primary description {primary_description} and secondary description {secondary_description} should be checked against index code {index_code}, and after reviewing the primary description {primary_description} again for consistency, it should be determined whether this crime is consistent with the assigned index code {index_code} according to the established classification criteria.';
\end{lstlisting}
\end{minipage}
\vspace{-5pt}
\caption{Q3 before Expression Compression}
\label{fig:q3:beforeer}
\vspace{4pt}
\end{figure}

\begin{figure}[H]
\centering
\begin{minipage}{0.45\textwidth}
\centering
\begin{lstlisting}{sql}
SELECT iucr_no FROM IUCR 
WHERE s'The index_code label rules are:
- index_code = I -> Indexed (severe, such as murder, rape, arson, and robbery)
- index_code = N -> Non-Indexed (less severe, such as vandalism, weapons violations, and peace disturbance)
Task: Determine whether the crime''s {primary_description} and {secondary_description} indicates this crime is consisted with {index_code}');
\end{lstlisting}
\end{minipage}
\vspace{-5pt}
\caption{Q3 before Partial Deduction}
\label{fig:q3:beforepd}
\vspace{4pt}
\end{figure}

\begin{figure}[H]
\centering
\begin{minipage}{0.45\textwidth}
\centering
\begin{lstlisting}{sql}
-- Note: Introduces repeated field references ({description} appears three times) and redundant sentence structures (repetitive "and the {description} indicates that..." patterns).
SELECT coins.id AS coin_id FROM coins
WHERE category = 'coin' AND status = 'active' AND description is NOT NULL
AND s'The {description} shows that this cryptocurrency has a big supply amount, and the {description} indicates that this cryptocurrency is still in circulation, and the {description} can be used to infer that the price of this cryptocurrency is rising.';
\end{lstlisting}
\end{minipage}
\vspace{-5pt}
\caption{Q6 before Expression Compression}
\label{fig:q6:beforeer}
\vspace{4pt}
\end{figure}

\begin{figure}[H]
\centering
\begin{minipage}{0.45\textwidth}
\centering
\begin{lstlisting}{sql}
SELECT coins.id AS coin_id FROM coins 
WHERE category = 'coin' AND description is NOT NULL
AND s'The {description} shows that this cryptocurrency has a big supply amount.'
AND s'The {description} indicates that this cryptocurrency is still in circulation.'
AND s'It can be inferred from the {description} that the price of this cryptocurrency is rising.'
\end{lstlisting}
\end{minipage}
\vspace{-5pt}
\caption{Q6 before Partial Deduction}
\label{fig:q6:beforepd}
\vspace{4pt}
\end{figure}

\begin{figure}[H]
\centering
\begin{minipage}{0.45\textwidth}
\centering
\begin{lstlisting}{sql}
-- Note: Adds redundant modifiers ("which describes the film narrative"), repeated field references ({overview} appears twice), and complex expressions ("after reviewing...for temporal clues").
SELECT movie_id FROM 
    (SELECT s'Based on the movie overview {overview} provided, which describes the film narrative, summarize the geographic and temporal background of the movie, including specific locations and time periods mentioned in this overview {overview}.' AS background, movie_id
        FROM movie
        SEMI JOIN (
            SELECT mg.movie_id
            FROM movie_genres mg JOIN genre g ON mg.genre_id = g.genre_id
            WHERE g.genre_name IN ('Thriller', 'Comedy', 'Western')
        ) mg ON m.movie_id = mg.movie_id
        WHERE revenue > budget
        AND length(overview) < 255
    )
    WHERE s'The background summary {background} indicates the story is set in modern times rather than historical periods, and after reviewing the background {background} for temporal clues, this setting classification is confirmed.';
\end{lstlisting}
\end{minipage}
\vspace{-5pt}
\caption{Q13 before Expression Compression}
\label{fig:q13-beforeer}
\vspace{4pt}
\end{figure}

\begin{figure}[H]
\centering
\begin{minipage}{0.45\textwidth}
\centering
\begin{lstlisting}{sql}
-- Note: Replaces originally SQL-expressible numerical comparison predicate revenue > budget with natural language description 'the {revenue} is greater than {budget}'.
SELECT movie_id FROM 
(SELECT s'Summarize the geographic and temporal background of the movie according to {overview}.' AS background, movie_id
    FROM movie
    SEMI JOIN (
        SELECT mg.movie_id
        FROM movie_genres mg JOIN genre g ON mg.genre_id = g.genre_id
        WHERE g.genre_name IN ('Thriller', 'Comedy', 'Western')
    ) mg ON m.movie_id = mg.movie_id
        WHERE s'the {revenue} is greater than {budget}'
        AND length(overview) < 255
)
WHERE s'{background} indicates the story is set in modern times rather than historical periods.';
\end{lstlisting}
\end{minipage}
\vspace{-5pt}
\caption{Q13 before Predicate Deduction}
\label{fig:q13:beforepd}
\vspace{4pt}
\end{figure}

\begin{figure}[H]
\centering
\begin{minipage}{0.45\textwidth}
\centering
\begin{lstlisting}{sql}
-- Note: Introduces lengthy classification descriptions and repeated field references ({plot} appears three times), adds unnecessary explanatory text.
SELECT plot,s'For genre classification purposes, the plot summary {plot} should be examined, and following a review of this plot {plot}, assignment to one of the predefined categories is required, including: SciFi, CrimeThriller, RomanceDrama, Horror, Comedy, War, Fantasy, FamilyAnimation, Biography, or if none match, classification as Other.' AS category 
FROM
(SELECT s'After reading the movie overview {overview}, which provides a description of the film, summarization of the main plot points and key narrative elements from this overview {overview} should be performed in a concise manner.' AS plot
    FROM movie m
    SEMI JOIN (
        SELECT ml.movie_id
        FROM movie_languages ml
        JOIN "language" l ON ml.language_id = l.language_id AND l.language_name = 'English'
    ) ml ON m.movie_id = ml.movie_id AND movie_status = 'Released' AND popularity > 10 AND length(overview) < 255 AND length(overview) > 0
);
\end{lstlisting}
\end{minipage}
\vspace{-5pt}
\caption{Q14 before Expression Compression}
\label{fig:q14:beforeer}
\vspace{4pt}
\end{figure}

\begin{figure}[H]
\centering
\begin{minipage}{0.45\textwidth}
\centering
\begin{lstlisting}{sql}
-- Note: Replaces SQL-expressible movie_status = 'Released' with originally natural language description '{movie_status} is released'.
SELECT plot,s'Classify {plot} into one of: Sci-Fi, Crime, Thriller, Romance, Drama, Horror, Comedy, War, Fantasy, Family, Animation, Biography, Other' AS category 
FROM
(SELECT s'Summarize the main plot of {overview}' AS plot
    FROM movie m
    SEMI JOIN (
        SELECT ml.movie_id
        FROM movie_languages ml
        JOIN "language" l ON ml.language_id = l.language_id AND l.language_name = 'English'
    ) ml ON m.movie_id = ml.movie_id AND s'{movie_status} is released' AND popularity > 10 AND length(overview) < 255 AND length(overview) > 0
);
\end{lstlisting}
\end{minipage}
\vspace{-5pt}
\caption{Q14 before Predicate Deduction}
\label{fig:q14-beforepd}
\vspace{4pt}
\end{figure}

\begin{tcblisting}{
  title=Prompt used for (partial) predicate deduction,
  colback=gray!5,
  colframe=gray!75,
  listing only,
}
You are a database optimizer assistant. Your task is to deduce the necessary SQL preconditions (pushdown predicates) for natural language expressions (NLEs), based on the provided column statistics.
An NLE is evaluated by an LLM to determine if a row satisfies a certain condition. Because LLM evaluation is expensive, we want to filter out rows that are guaranteed to be false by executing simple SQL predicates on the CPU first.
IMPORTANT: Your goal is NOT to convert the NLE into an equivalent SQL statement. Instead, you must find the necessary conditions of the NLE predicate given the column statistics.
The provided statistics for each column include:
- whether it is nullable
- the number of distinct values
- the top-5 most frequent values (sampled from a subset of rows)
Steps:
1. Understand the semantic meaning of the NLE.
2. Examine the column statistics, especially the top frequent values.
3. Identify values that definitively violate the NLE predicate.
   For instance, nan, N/A, empty strings, or NULL often represent missing data.
   If the NLE predicate requires a valid or meaningful value, these missing values can be filtered out.
4. Formulate DuckDB SQL predicates that filter out these violating values.
If you can deduce necessary conditions, output them as a valid JSON array of strings.
Each string must be a valid DuckDB SQL predicate (e.g., "col != 'nan'").
The final filter will be the logical AND of all strings in the array.
If no meaningful necessary condition can be deduced, simply output an empty array.
Example:
NLE Predicate:
Translated\_review is a positive user review
Column stats:
Translated\_Review: nullable=true, distinct=10537,
top5=["nan":711, "Good":13, "Negative":10, "Great":6, "Really good":5]
Output:
["Translated\_Review != 'nan'", "Translated\_Review != 'Negative'"]
Note:
The output MUST be a valid JSON array of strings (or an empty array).
\end{tcblisting}

\begin{tcblisting}{
  title=Prompt used for (partial) predicate deduction verification,
  colback=gray!5,
  colframe=gray!75,
  listing only,
}
You are a database optimizer assistant. Your task is to verify whether each candidate SQL predicate is a NECESSARY CONDITION for a set of Natural Language Expressions (NLEs).
Background:
A semantic filter evaluates NLEs (natural language conditions) on each row. Because LLM evaluation is expensive, we pre-filter rows using cheap SQL predicates. A SQL predicate P is a valid pushdown filter only if it is a NECESSARY CONDITION: whenever P is false, the overall NLE filter is guaranteed to be false.
KEY CONCEPT - Necessary vs Sufficient:
For each column, some values make NLE FALSE (set A), some make it TRUE (set B).
- **Necessary (valid)**: \"column != value\" for values in A. Filtering out violating values is correct.
- **Sufficient (invalid)**: \"column = value\" or \"column IN (...)\" for values in B. Whitelisting known-good values is WRONG: it would exclude valid rows whose values are not in the statistics. Output false for such predicates.
Definition of necessary condition:
  SQL predicate P is a necessary condition for NLE set S iff:
  for every possible row, NOT P => (at least one NLE in S is false)
  Equivalently: if a row passes all NLEs, it must also satisfy P.
Given the original NL Expressions and the candidate SQL Predicates, output a JSON array of booleans (same length as the SQL Predicates list):
 - true: the SQL predicate IS a valid necessary condition (typically \"column != value\" filtering out violating values)
  - false: the SQL predicate is NOT valid (too strict, unrelated to NLEs, or a sufficient condition like \"column = value\" / \"column IN (...)\")
Example:\n"
NL Expressions: [\"{Review} is a positive and meaningful user review\"]
SQL Predicates: [\"Review != 'nan'\", \"Review = 'Good'\", \"Rating > 10\"]
Output: [true, false, false]
(Review != 'nan' is necessary: 'nan' violates the NLE. Review = 'Good' is NOT necessary: it is a sufficient condition that would wrongly exclude other positive reviews. Rating > 10 is unrelated to the NLE.)
Note: Output MUST be a valid JSON array of booleans only.;

\end{tcblisting}

\end{document}